\documentclass[prx,showpacs,twocolumn,superscriptaddress,floatfix]{revtex4-1}

\usepackage{epsfig}
\usepackage{amsmath}
\usepackage{amssymb}
\usepackage{amsfonts}
\usepackage{mathptmx}
\usepackage{dcolumn}
\usepackage{eucal}
\usepackage{bm}
\usepackage{color}
\usepackage[colorlinks,linkcolor=blue,citecolor=blue]{hyperref}
\usepackage{epstopdf}
\usepackage{url}

\usepackage{dsfont}

\def\eps{\varepsilon}

\def\bk{{\bf k}}

\def\up{\uparrow}
\def\dw{\downarrow}

\def\be{\begin{equation}}
\def\ee{\end{equation}}
\def\bea{\begin{eqnarray}}
\def\eea{\end{eqnarray}}
\def\ba{\begin{array}{l l}}
\def\ea{\end{array}}
\begin{document}

\title{Spectroscopic Signatures of Gate-Controlled Superconducting Phases}

\author{Maria Teresa Mercaldo}
\affiliation{Dipartimento di Fisica ``E. R. Caianiello", Universit\`a di Salerno, IT-84084 Fisciano (SA), Italy}

\author{Francesco Giazotto}
\affiliation{NEST, Istituto Nanoscienze-CNR and Scuola Normale Superiore, Piazza San Silvestro 12, I-56127 Pisa, Italy}

\author{Mario Cuoco}
\affiliation{SPIN-CNR, IT-84084 Fisciano (SA), Italy}
\affiliation{Dipartimento di Fisica ``E. R. Caianiello", Universit\`a di Salerno, IT-84084 Fisciano (SA), Italy}
%\pacs{pacs}

\begin{abstract}
We investigate the tunneling conductance of superconductor-insulator-normal metal (SIN) and superconductor-insulator-superconductor (SIS) heterostructures with one superconducting side of the junction that is electrically driven and can exhibit $\pi$-pairing through a modification of the surface inversion asymmetric couplings. In SIN tunneling we find that the variation of the electrically driven interactions generally brings an increase of quasi-particles in the gap due to orbitally polarized depaired states, irrespective of the inter-band phase rearrangement. The peak of SIN conductance at the gap edge varies with a trend that depends both on the strength of the surface interactions as well as on the character of the gate-induced superconducting state. While this shift can be also associated with thermal effects in the SIN configuration, for the SIS geometry at low temperature the electric field does not yield the characteristic matching peak at voltages related with the difference between the gaps of the superconducting electrodes. This observation sets out a distinctive mark for spectroscopically distinguishing thermal population effects from signatures that are mainly related to a variation of the electric field. In SIS the electrostatic gating yields a variety of features with asymmetric peaks and broadening of the conductance spectral weight. These findings indicate general qualitative trends for both SIN and SIS  tunneling spectroscopy which could serve for evaluating the impact of electric field on superconductors and the occurrence of non-centrosymmetric orbital antiphase pairing.       
\end{abstract}
\maketitle

\section{Introduction}

Understanding the interplay of electricity and superconductivity is a fundamental problem that stands out for its general relevance, the great impact it can have for accessing, controlling or driving new phases of quantum matter, and the enticing  perspectives for the development of future quantum technologies. Electric field effects have been successfully employed to drive or control the superconducting phase in materials with low-to-moderate carrier density \cite{Ueno2008,Ueno2011,Ye2012,Bollinger2011}, in thin films \cite{Shi2015,Leng2012,Shiogai2016} and interfaces \cite{Caviglia2008}, down to the two-dimensional limit \cite{Li2016,Saito2015,Chen2019}. 
In this context, to achieve a control of the charge-carrier density by gating the materials have to be in a low-density regime and thin enough to avoid shielding of the electric field. In fact, for superconductors with large electron density the strong electric screening typically allows small changes in the superconducting critical temperature \cite{Choi2014,Piatti2017,Ummarino2017,Piatti2016}. 

Recently, the discovery of unconventional gating effects in metallic superconductors \cite{Paolucci2019a} has challenged the current view of how electric field can affect superconductivity. 
In particular, the overall experimental observations which have been so far accumulated, provide a variety of unexpected fingerprints of the consequences of electrostatic gating on superconductors. The major findings demonstrate the following effects: reduction or full suppression of the critical supercurrent in nanowires \cite{DeSimoni2018,Ritter2020,Alegria2020} and Dayem bridges \cite{Likharev1979,Paolucci2018,Paolucci2019b,Puglia2020}, manipulation of superconducting phase in interferometric setups \cite{Paolucci2019b,Paolucci2019c}, enhancement of phase fluctuations \cite{Puglia2020a}, increase of in-gap quasiparticle population \cite{Ritter2020}, and weak interrelation between the critical magnetic field and the critical voltage associated with the vanishing supercurrent transition \cite{Bours2020}. Remarkably, a suppression of the critical supercurrent has been also achieved in fully suspended gate-controlled Ti nano-transistors \cite{Rocci2020} and with ionic-gating \cite{Paolucci2021}, thus posing bounds on the role of charge injections or electron leakage within the observed phenomena. 

The emerging experimental scenario is definitely opening fundamental challenges about the coupling of static electric field and superconductivity. 
Indeed, the impact of electrostatic gating on the metallic superconductor points to different channels of interaction. One possibility is that the application of large electrostatic gating may cause the injection of highly energetic quasiparticles. In this framework, the induced excitations in the superconductor, with energies that are greater than the superconducting gap, would mainly lead to thermally driven population unbalances and non-equilibrium effects which substantially may end up into a suppression of the amplitude of the superconducting order parameter. Along this line, the increase of the quasiparticle in-gap spectral weight in the tunneling conductance has been indeed recently interpreted as an evidence of high-energy injection of quasiparticles into the superconductor \cite{Alegria2020,Ritter2020}.

\begin{figure*}[bt]
\includegraphics[width=0.35\textwidth]{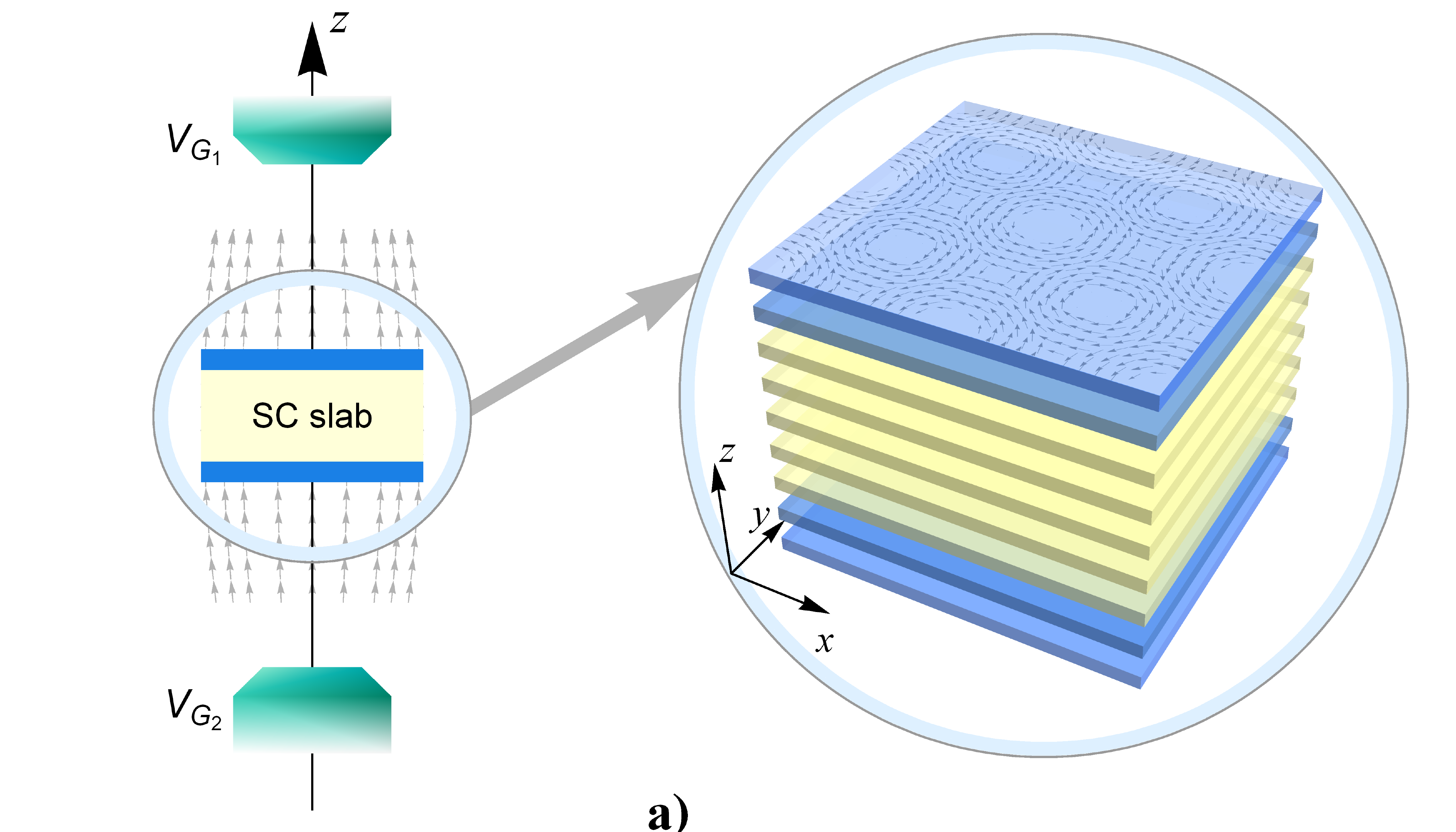} 
\hspace{0.04cm}
\includegraphics[width=0.38\textwidth]{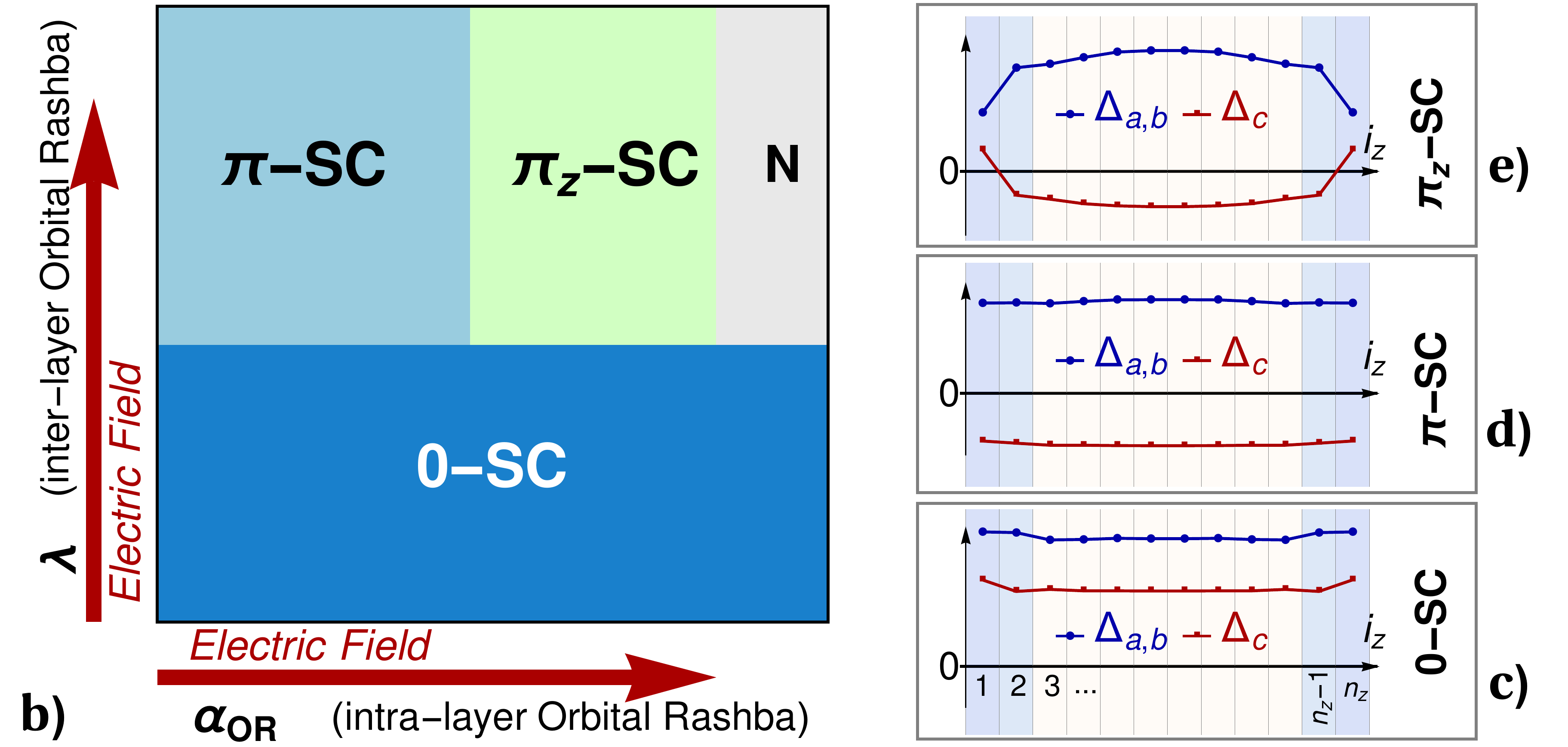}
\hspace{0.04cm}
\includegraphics[width=0.22\textwidth]{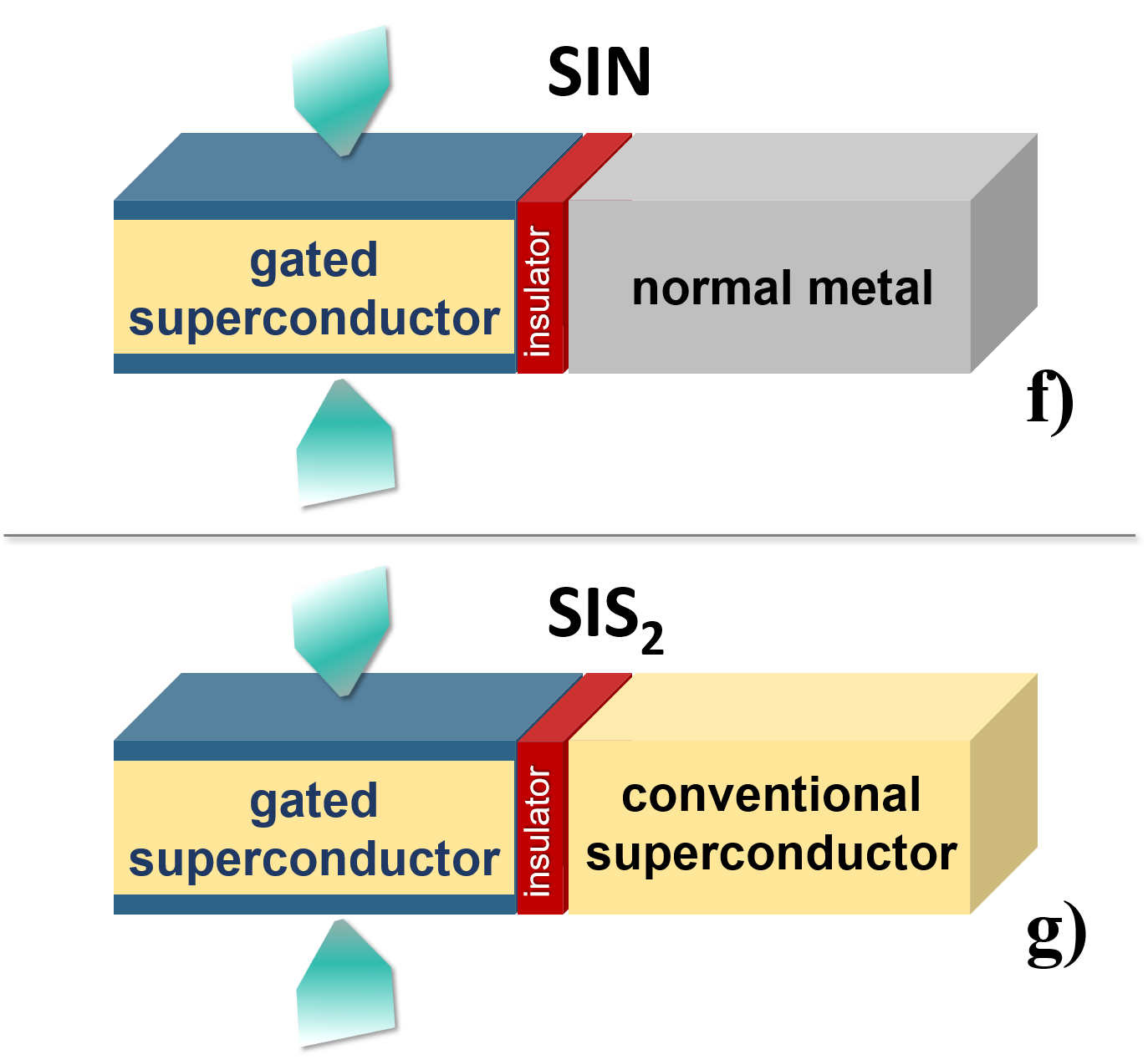}
\protect\caption{(a) Illustration of the superconducting slab embedded into an electrostatic field generated by gates with voltages $V_{G_1}$ and $V_{G_2}$. For the multilayered superconducting thin-film, due to the screening on the Thomas-Fermi atomic length, one can distinguish between the surface layers (blue), where the electric field fully penetrates and affects the electronic structure, from the inner layers (yellow) that are not directly influenced by electric field induced interactions. On the top surface layer we show a representative pattern of the chiral textures in the Brillouin zone due to the electric field induced orbital-Rashba couplings.
(b) Schematic illustration of the superconducting phase diagram which is obtained by applying an electric field through the modification of the induced intra- ($\alpha_{\mathrm{OR}}$) and inter- ($\lambda$) orbital polarizing interactions at the surface layers of the superconducting thin film. Within the phase diagram, there occurs a conventional spin-singlet superconducting state with orbitally polarized surface states, that is labeled as 0-SC (c). Otherwise, the electric field generates $\pi$-pairing with the superconducting order parameter having a $\pi$-phase shift among different bands at the Fermi level. Here, we consider a three-band modeling with $\Delta_a$,$\Delta_b$, and $\Delta_c$ the corresponding spin singlet order parameters. 
$\pi_z-$SC stands for an inhomogeneous superconducting phase along the confining direction of the thin film ($z$) with layers having zero interband phase shift alternating with layers marked by $\pi$-pairing (e). Finally, $N$ indicates the normal metal state configuration with strong orbitally polarized depaired states at the Fermi level. An overall schematic illustration of the $0$-, $\pi$- and  $\pi_z$-SC phases is reported in (c-e) with the layer and band dependent superconducting order parameters determined by iterative computation on the lowest energy state. Schematic of the SIN (f) and SIS (g) junction with one side subjected to electrostatic gating. The coloured surface on the gated superconductor sketches the penetration of the electrostatic field.}
\label{f1}
\end{figure*}

Another scenario places its roots in the fundamental interaction between an electrostatic potential and the electronic states on the surface of the superconductor through the modification of the strength of the inversion asymmetric interaction \cite{Mercaldo2020}. Electric field breaks inversion symmetry on the surface and can lead to orbital polarization of the electronic states through orbital Rashba couplings that, however, are active only in the spatial region where the electrostatic field can penetrate the metal, i.e. of the order of the Thomas-Fermi length \cite{Virtanen2019}. 
In this context, it has been indeed recently recognized that an analog of the spin Rashba coupling \cite{Rashba1960, Dresselhaus1955} at the surface of low-dimensional electron systems or non-centrosymmetric materials arises due to the coupling between the atomic orbital angular momentum ${\bf L}$ and the crystal wave-vector ${\bf k}$ \cite{Park2011,Park2012,Kim2013,Mercaldo2020}. 
For materials having configurations close to the Fermi level with non-vanishing atomic angular momentum, an intrinsic crystalline potential or an applied electric field, breaking spatial inversion or mirror, yield non-local odd-parity matrix elements among atomic orbitals contributing to the Bloch states. 
The resulting orbital Rashba effect \cite{Park2011,Park2012,Kim2013,Mercaldo2020}, in analogy with the spin Rashba effect, refers to an orbital angular momentum dependent energy splitting in the Brillouin zone that yields chiral orbital texture \cite{Park2013,Kim2012,Kim2013,fukaya19,Mercaldo2020}. The orbital Rashba effects and the resulting orbitally polarized textures have been investigated in various materials, including elemental noble metal surfaces \cite{Kim2012}, surface alloys \cite{Go2017,Unzelmann2020}, topological surface states \cite{Park2012-1}, bulk ferroelectrics \cite{Ponet2018}, oxide surface \cite{Sunko2017} and interfaces.

The essential consequence of such orbital polarizing effects is that the electric field is able to break the inter-band superconducting phase rigidity or fully suppress the amplitude of the order parameter \cite{Mercaldo2020}. 
%The breaking of the inter-orbital phase rigidity is a consequence of the fact that the electric field violates inversion symmetry within the surface layers and thus orbitally polarizes the electronic states at the Fermi level. 
Then, two main consequences are in place as schematically depicted in the phase diagram of Fig. \ref{f1}. The gating can rearrange the band dependent superconducting phases with a $\pi$-shift (i.e. resulting into a $\pi$- or $\pi_z$- phase depending on the spatial homogeneity of the $\pi$-pairing, Fig. \ref{f1} (d)-(e)), or it can completely suppress the amplitude of the order parameter by increasing the population of depaired orbitally polarized quasi-particles until reaching the normal phase $N$. Both phases (i.e. the $\pi$-states and the normal metal with strong orbitally polarized Fermi surfaces) can be in principle marked by a vanishing supercurrent, however the underlying mechanism that leads to the supercurrent suppression is fundamentally different. In fact, for the $\pi$-states the supercurrent suppression is due to the superconducting phase frustration of different bands (i.e. it mainly arises from the interference of the phases of the superconducting order parameters at the Fermi level), while for the electrically induced normal metal phase it is essentially due to a suppression of the amplitude of the superconducting order parameter. Remarkably, the application of a magnetic field can help to discern between the two scenarios concerning the way the electric field impacts on the amplitude and phase of the superconductor \cite{Bours2020}. In this context, the weak dependence of the critical magnetic fields on the critical voltages and viceversa seems to be mostly compatible with a scenario with the electrostatic gating driving the transition from a conventional superconducting phase to a superconducting state with $\pi$-pairing \cite{Mercaldo2020} where the inter-band phase frustration is responsible of the vanishing of the supercurrent.

It is interesting to point out that $\pi$-pairing is also expected to occur in unconventional superconductors in the absence of external perturbations as indicated, for instance, by the recent observations in iron-based materials \cite{Grinenko2020} or theoretically proposed at oxides interface \cite{Fukaya2020,Scheurer2015}. Hence, the addressed problem has a wider framework and our results can have a more general application. Indeed, they can provide insight into the tunneling spectroscopy of superconducting thin films where the inter-band phase reconstruction already manifests in the absence of an applied electric field and the application of other drives can affect the orbital polarization of the electronic state on the surface and further reconstruct the superconducting phase. 

Starting from this physical outlook it is relevant to consider whether spectroscopically one can assess the nature of the gate-driven superconducting phases and search for fingerprints which can be employed to understand the way the electric field affects the superconductivity. For this purpose, we explore the tunneling conductance of superconductor-insulator-normal metal (SIN) and superconductor-insulator-superconductor (SIS) heterostructures with one superconducting side of the junction being subjected to the electrostatic gating. 
Our strategy is to evaluate the tunneling conductance for SIN and SIS configurations in the phase space spanned by the electric field through the variation of the interactions at the surface layers of the superconductor moving within different types of $\pi$-paired configurations. 
The key target is to track the evolution of the tunneling conductance in all phases that are obtained within the scenario of an electrostatically triggered orbital polarization at the surface (Fig. \ref{f1}(b)) and extract the main spectral signatures. In this way one can get a significant insight into the spectroscopic response that can result when considering the gate-driven superconducting transitions. Starting from the SIN tunneling spectroscopy the electric field is shown to yield an increase of spectral weight in the gap, due to the orbitally polarized depaired states, and this trend is observed independently of the allowed inter-band phase rearrangements. The voltage position of conductance peak nearby the gap edge varies with a tendency that is sensitive to the amplitude of the orbital Rashba interactions as well as the character of the gate-driven superconducting state. While this shift can be also associated with thermal effects, for the SIS geometry at low temperature the electric field does not yield the characteristic matching peak for voltages that correspond to the difference between the gaps of the superconductors. This observation thus represents a distinctive mark for spectroscopically disentangling the thermal population effects from signatures that are mainly related to a variation of the electric field. Apart from such feature, in SIS the electric field yields a variety of marks with asymmetric redistribution of the spectral weight of the main conductance peak. These characteristics when tracked as a function of the electric field indicate distinct trends for both SIN and SIS configurations with characteristic tunneling behavior in gate-controlled superconducting phases. 

The paper is organized as follows. In Sect. II we present the model and the methodology for the tunneling conductance analysis. Sect. III is devoted to the main results by focusing on the SIN and SIS tunneling spectroscopy. In Sect. IV we provide the discussion on the resulting effects and the concluding remarks. In the Appendix we provide details of the profile of the electrostatic potential close to the surface of the superconductor. Additionally, we present the basic elements for the derivation of the tight-binding model in the presence of an electrostatic potential at the surface and the temperature dependence of orbital dependent superconducting order parameters obtained by self-consistently solving the gap equations.

\section{Model and methodology}

In this section we present the model and the methodogy which have been employed in order to evaluate the tunneling conductance for the examined SIN and SIS geometries. To achieve the goal, we need determine the density of states of the superconducting side of the junction that is subjected to the electrostatic gating. For this part of the heterostructure, the superconductor is considered to have a slab geometry with $n_z$ layers (Fig. \ref{f1} (a)) and we assume a conventional $s-$wave spin-singlet pairing. 
%Then, to further include the orbital polarizing effects, as pointed out in the introduction, we employ an electronic description of the gate-controlled superconductor that is based on anisotropic $d$-orbitals since they can develop a non vanishing atomic orbital angular momentum. For instance one can focus on the ($yz,xz,xy$) subsector of the $d$-shell, but similar modelling can also apply to $p$- and $sp$-bands \cite{Mercaldo2020}. 

Before going into the details of the results for the SIN and SIS conductance we deepen the discussion about the main aspects of the model that has been used to describe the effects of the applied electric field. 
Concerning the inclusion of the gate voltage in the employed model Hamiltonian, there are three relevant working hypotheses.
\\
Firstly, we are considering a metallic superconductor (i.e. with a high density of charge carriers and thus large Fermi surface). In this context, the electrostatic potential in the inner layers of the superconductor is vanishing due to screening effects for distances from the surface that are greater than the Thomas-Fermi length $\lambda_{\text{TF}}$ (typically of few unit cells along the out-of-plane $\hat{z}$-direction).
The second aspect concerns the character of the gate voltage. We assume to insert the layered superconducting system in a capacitor like structure (open circuit configuration), that fixes the value of the gate voltage to be opposite at the two sides of the layered superconductor (see Fig. 1(a)).
Then, a final point refers to the investigated electrical regime. We deal with stationary conditions, i.e. constant in time electrostatic gating. Hence, the reference Maxwell equation to be considered is that one for the scalar electrostatic potential. Basically, within the capacitor configuration we have $\nabla^2 V(r)=0$ in the region between the plates and the surface of the superconductor, while $\nabla^2 V(r)-k^2_{\text{TF}} V(r)=\rho(r)/(4 \pi \epsilon_0)$ inside the superconductor, with $k_{TF}=\lambda_{\text{TF}}^{-1}$ being the inverse of Thomas-Fermi length, and $\rho(r)$ the induced charge at the surface.

Taking into account the above assumptions, one can determine the profile of the electrostatic potential nearby the surface of the superconductor. 
%assuming that a voltage difference is applied at the plates of the capacitor. 
Since the screening effects in a metal leads to an exponential decaying of the electric field within a distance of the order of few unit cells, the effects of the electrostatic potential $V(z)$ are confined at the surface layers of the slab. 
Hence, without loss of generality and due to the geometry (i.e. the thin film has a dimension in the $xy$ plane that is larger than the thickness along the $z$-direction) and symmetry of the problem, the electrostatic potential can be taken as only dependent on the $z$ coordinate and exponentially vanishing inside the superconductor. The solution of the electrostatic equation for this configuration can be handled  and is reported in Appendix A. 
To proceed further and construct the tight-binding model for the layered superconductor, one has to evaluate the matrix elements of $V(z)$ for the selected multi-orbitals Wannier basis having non-vanishing atomic angular momentum $L$. For this purpose, we take a 3$\times$3 sector which is suitable for $p-$ or $d-$ (e.g. $t_{2g}$ multiplet of the $d$-manifold in cubic/tetragonal symmetry) states with $L=1$, being a description that can be effectively applied to a large class of materials. In this manifold the electrostatic potential has diagonal and off-diagonal elements. The diagonal terms renormalize the chemical potential while the off-diagonal ones lead to the orbital Rashba coupling. 
In the employed model, we are neglecting the modification of the chemical potential at the surface layer because the correction to the electron density is substantially affecting only the surface layer and is negligible for a metallic system (see Appendix A), with also a minor impact on the superconducting order parameter \cite{Virtanen2019}.
In turn, the off-diagonal terms lead to orbital Rashba couplings. In Appendix B we provide the main steps for the derivation of the orbital Rashba interaction at the surface.
%has been obtained by taking into account the leading linear order term of $V(z)$ close to the surface.

Hence, since the applied electric field on the surface of the superconductor is parallel to the out-of-plane $\hat{z}$-direction and thus the electrostatic potential close to the surface, at the linear order in $z$, can be described by a potential $V_{s}=-E_{s} z$ with $E_s$ being constant in amplitude (assuming the electric charge $e$ is unit). Taking into account the steps for the derivation of the tight binding model (see Appendix B) \cite{Park2011,Park2012,Kim2013,Mercaldo2020,Bours2020}, the matrix elements of $V_s$ in the Bloch basis can yield an intra- ($\alpha_{OR}\sim E_s$) and inter-layer ($\lambda \sim E_s$) inversion asymmetric interactions, whose ratio depends on the inter-atomic distances and distortions occurring at the surface layers \cite{Mercaldo2020}.
For convenience one can indicate as $(a,b,c)$ the ($yz,xz,xy$) $d-$orbitals. Then, after introducing the creation $d^\dagger_{\alpha,\sigma}(\bk,i_z)$ and annihilation $d_{\alpha,\sigma}(\bk,i_z)$ operators with momentum $\bk$, spin ($\sigma=[\up,\dw]$), orbital ($\alpha=(a,b,c$)), and layer $i_z$, one can construct a spinorial basis
$\Psi^\dagger(\bk,i_z)=(\Psi_{\up}^\dagger(\bk,i_z), \Psi_{\dw}(-\bk,i_z))$ with
$\Psi_{\sigma}^\dagger(\bk,i_z)=(d^\dagger_{a,\sigma}(\bk,i_z),d^\dagger_{b,\sigma}(\bk,i_z),d^\dagger_{c,\sigma}(\bk,i_z))$. In this representation, the Hamiltonian can be generally expressed in a compact way as \cite{Mercaldo2020,Bours2020}:
\begin{eqnarray}
\mathcal{H}= \frac{1}{N} \sum_{\bk,i_z,j_z} \Psi^{\dagger}(\bk,i_z) \hat{H}(\bk) \Psi(\bk,j_z) \,,
\end{eqnarray}
\noindent with 
\begin{eqnarray}
&& \hat{H}(\bk)=\sum_{\alpha=\{a,b,c\}} [\tau_z \eps_\alpha(\bk) + \Delta_\alpha(i_z) \tau_x ]\otimes(\hat{L}^2 - 2 \hat{L}^2_\alpha)] \delta_{i_z,j_z} + \nonumber \\
&&+ \alpha_{OR} \tau_z \otimes (\sin k_y \hat{L}_x - \sin k_x \hat{L}_y) [\delta_{i_z,j_z}(\delta_{i_z,1} + \delta_{i_z,n_z})] + \nonumber \\
&& + t_{\perp,\alpha} \tau_z \otimes (\hat{L}^2 - 2 \hat{L}^2_\alpha) \delta_{i_z,j_z\pm 1}+\nonumber \\
&& +  \lambda \left[(\hat{L}_x+\hat{L}_y) (\delta_{i_z,1}\delta_{j_z,2}+\delta_{i_z,n_z}\delta_{j_z,n_z-1}) +\text{h.c.} \right] \,,
\end{eqnarray}
\noindent where the orbital angular momentum operators $\hat{L}$ have components 
$
\hat{L}_x=\begin{bmatrix}
0 & 0 & 0 \\ 
0 & 0 & i \\ 
0 & -i & 0%
\end{bmatrix}, 
\hat{L}_y=\begin{bmatrix}
0 & 0 & -i \\ 
0 & 0 & 0 \\ 
i & 0 & 0%
\end{bmatrix}, \hat{L}_z=\begin{bmatrix}
0 & -i & 0 \\ 
i & 0 & 0 \\ 
0 & 0 & 0%
\end{bmatrix}$
within the ($yz,xz,xy$) subspace, $\tau_i$ ($i=x,y,z$) are the Pauli matrices for the electron-hole sector, and $\delta_{i,j}$ the Kronecker delta function. 
Due to the anisotropy and directionality of the $d$-orbitals, the kinetic energy for the in-plane electron itinerancy is expressed by $\eps_a(\bk)=-2 t_{||} [\eta \cos(k_x)+\cos(k_y)]$, $\eps_b(\bk)=-2 t_{||} [\cos(k_x)+ \eta \cos(k_y)]$, and $\eps_c(\bk)=-2 t_{||} [\cos(k_x)+ \cos(k_y)]$, with $\eta$ being a term that takes into account deviations from the ideal cubic symmetry. Other long-range terms or processes involving inter-orbital hoppings that are activated by distortions do not change the quality of the addressed phenomenology \cite{Mercaldo2020}.
We assume that the layer dependent spin-singlet order parameter (OP) is non-vanishing only for electrons belonging to the same band. 
{{Here, we have neglected the pair-hopping term of the form $V_{\alpha\beta}(k,i_z)=J_{\alpha\beta} d^{\dagger}_{\alpha,\uparrow}(k,i_z) d^{\dagger}_{\alpha,\downarrow}(-k,i_z ) d_{\beta,\uparrow}(k,i_z) d_{\beta,\downarrow}(-k,i_z )$. In the early works on two-band superconductivity \cite{Suhl1959,Kondo1963,Leggett1966}, the pair-hopping term has been included to remove the degeneracy among the configurations with different phase difference between the intra-orbital superconducting order parameters in systems without single particle orbital hybridization. For the examined model, we have inter-orbital mixing and the 0- and $\pi$- phases are separated in energy (see for instance Fig. 3 in Ref. \cite{Mercaldo2020}), with the 0- configuration being the ground-state in absence of an applied electric field even without the pair hopping $V_{\alpha\beta}$ terms. 
%This is because the single-particle inter-orbital mixing tends to stabilize a superconducting ground state with zero inter-band phase difference. 
In this respect, the inclusion of the term $V_{\alpha\beta}$ would further stabilize the 0-state in the phase diagram and one would need a slightly larger inter-layer orbital Rashba coupling to induce the transition from the 0- to the $\pi$-phase. Thus, since for the study of the tunneling conductance we have investigated all the regimes in the phase diagram, we do not expect qualitative changes in the results.}}

%and it is expressed as $\Delta_\alpha(i_z)=\frac{1}{N} \sum_{\bk} g\,\langle d_{\alpha,\uparrow}(\bk,i_z) d_{\alpha,\downarrow}(-\bk,i_z)  \rangle$ with $\langle ...\rangle$ being the expectation value on the ground state. 
Concerning the evaluation of the superconducting order parameters, $\Delta_\alpha(i_z)=\frac{1}{N} \sum_{k} g\,\langle d_{\alpha,\uparrow}(k,i_z) d_{\alpha,\downarrow}(-k,i_z)  \rangle$, we performed it by computing the trace of the pairing operator $\hat{P}^{\alpha,i_z}_{k} =d_{\alpha,\uparrow}(k,i_z) d_{\alpha,\downarrow}(-k,i_z)$ over all the eigenstates $|n,k\rangle$ of the Hamiltonian associated to negative energies $E_{n,k}<0$ at zero temperature (or to all energy configurations at finite temperature weighted by the Fermi function). Since the eigenstates $|n,k\rangle$ depend on $\Delta_\alpha(i_z)$ and the orbital Rashba interactions couple the momentum with the local angular momentum, the gap equations of the orbital dependent order parameters are strongly coupled between each other. To make evident this dependence one can express the gap equation in the following way
\begin{eqnarray*}
\Delta_\beta (i_z) &=&
\frac{1}{N} \sum_{k,n} g\,\langle n,k;\Delta_{\alpha}(i_z)| d_{\beta,\uparrow}(k,i_z)  \\ 
&\times& d_{\beta,\downarrow}(-k,i_z)  |n,k;\Delta_{\alpha}(i_z)\rangle f(E_{n,k},T) 
\end{eqnarray*}
\noindent with $f(E_{n,k},T)=\frac{1}{\exp(E_{n,k}/k_B T)+1}$ the Fermi-Dirac distribution at given temperature $T$.
In this framework, we have computed self-consistently the order parameters for each orbital character as a function of the temperature (see Appendix B). Since the deviation from the canonical BCS behavior does not alter the qualitative outcome of the analysis, in order to have a uniform comparison of the various regimes, we have determined the conductances by taking the BCS profile for the temperature dependence.

Here, $N=n_x \times n_y$ sets the dimension of the layer in terms of the linear lengths $n_x$ and $n_y$, while translation invariance is taken in the $xy$-plane and $n_z$ is the number of layers along the $z-$axis (Fig. \ref{f1}). We notice that $g$ is not modified by the electric field. This is physically plausible because due to screening effects the electric field cannot induce an inversion asymmetric potential inside the thin film on distances from the surface that goes beyond the Thomas-Fermi length.

{{For completeness, we point out that the ground state of the investigated model Hamiltonian cannot sustain a time dependent phase dynamics under the applied voltage. 
In order to get a non-trivial time dependence of the Josephson phase it is crucial to have two superconducting condensates separated by an insulator that is thin enough to keep an energy difference of the Cooper pairs as given by the applied voltage. In turn, for the examined superconducting multi-layered system, we have only one superconducting condensate. The voltage difference across the surface layers does not lead to an energy difference among the Cooper pairs in neighbor layers as they are not isolated and they are in good electrical contact since we are dealing with a metallic system. Although there is an inhomogeneous electrostatic potential at the surface compared to the inner layers, both the single particle states and the Cooper pairs are not localized in the corresponding regions of the superconductor. Additionally, there is no phase difference between the superconducting order parameters at the top and bottom surface layers (or among those at the surface layers and the neighbor inner layers) because such phase difference would be associated to a non-equilibrium state with a current flowing across the slab. Such flow would then lead to charge imbalance across the superconductor which is energetically unfavourable and prevented by the electrical screening in the metallic superconductor. According to these observations, it is useful to remark that this regime for the phase in the superconducting system can be broken if the superconducting materials are marked by inhomogeneities in the region where the electric field penetrates that yield isolated islands or clusters separated by insulating barriers, i.e. spatially distributed effective weak-links near the surface of the superconductor. 
}}
\\
Concerning the model Hamiltonian in Eq. 2, we also observe that it has only one term that is not compatible with the charge conjugation transformation involved in the time reversal symmetry operation. This term is the inter-layer orbital Rashba interaction which changes sign under a complex conjugation transformation. Such aspect, however, has an impact only on the phase of the superconducting order parameter. This consequence has been taken fully into account by allowing a complex value for the orbital-dependent order parameters in the self-consistent simulation of the superconducting order parameter. 
Moreover, since this is an inter-layer orbital dependent charge transfer that is activated by the electrostatic gating only at the surface of the superconductor, it does not affect the structure of the s-wave pairing in the inner layers which keeps its form of zero momentum with electron pairs having opposite spin at $k$ and $-k$ close to the Fermi surface. 
We observe that the electric field is a surface perturbation and does not influence at all the pairing interaction in the inner layers of the superconductor, thus the employed conventional form of the superconducting order parameter is a physically valid assumption for the examined model.  
For completeness, we also point out that a kind of pair density wave along the $z-$ direction (i.e. $\pi_z$ phase) can be achieved in the strong coupling regime of orbital Rashba interactions \cite{Mercaldo2020}. The occurrence of this type of pair density reflects the fact that, due to the electrostatic potential, a reconstruction of the order parameter can occur along the $z$- direction. This outcome does not imply a variation in the structure of the in-plane pairing.

%%%%%
\begin{figure}[t!]
\includegraphics[width=0.8\columnwidth]{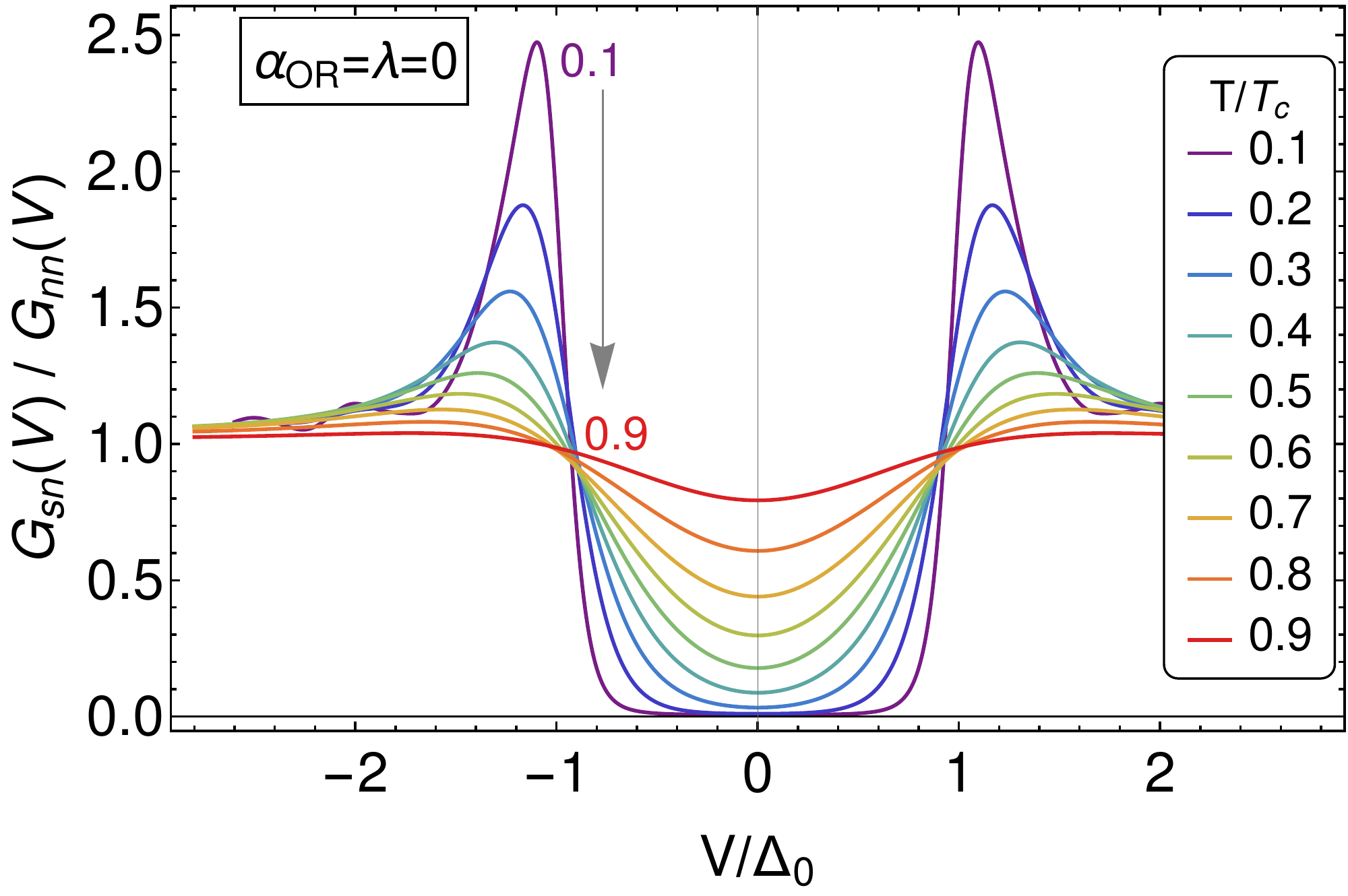}
\protect\caption{Thermal evolution of the normalized tunneling conductance, $G_{sn}(V)$, for the SIN junction evaluated in the configuration with vanishing external electrostatic gating (i.e. for $\alpha_{\text{OR}}=\lambda=0$ within the implemented modeling) for several temperatures $T$ below the superconducting critical temperature $T_c$. $\Delta_0$ is is the amplitude of the lowest energy excitation for the zero electric field state assuming a uniform profile over the layers of the superconducting order parameter. The computation has been performed for a superconducting electrode with $n_z=6$ layers. For the temperature dependence of the superconducting order parameter we employ the phenomenological BCS expression $ \Delta(T)=\Delta_0 \sqrt{\cos(\pi T / 2 T_c)}$ \cite{Sheahen1966}.}
\label{f2}
\end{figure}
%%%%

%%%%%
\begin{figure}[bt]
\includegraphics[width=0.48\columnwidth]{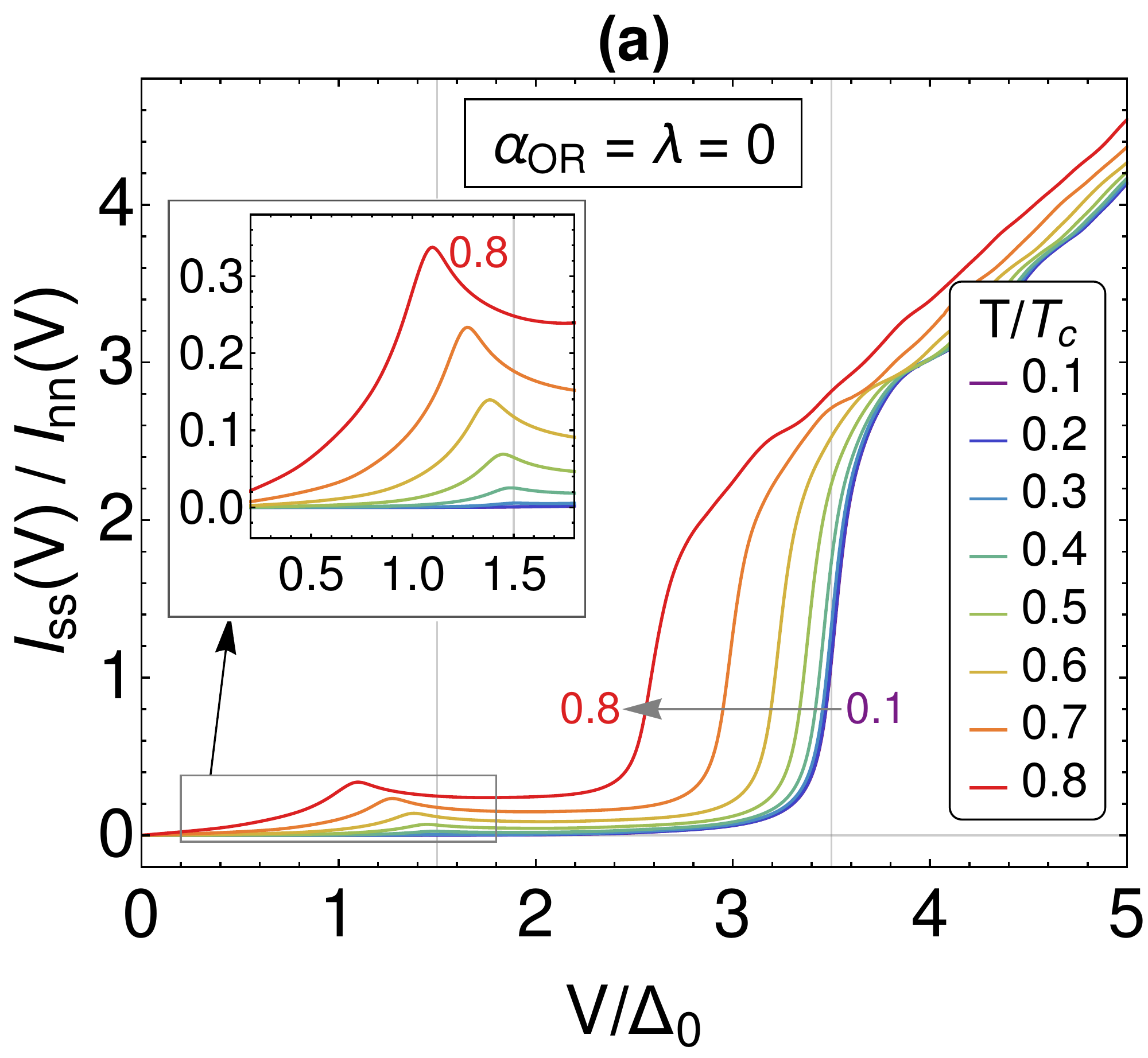}
\includegraphics[width=0.48\columnwidth]{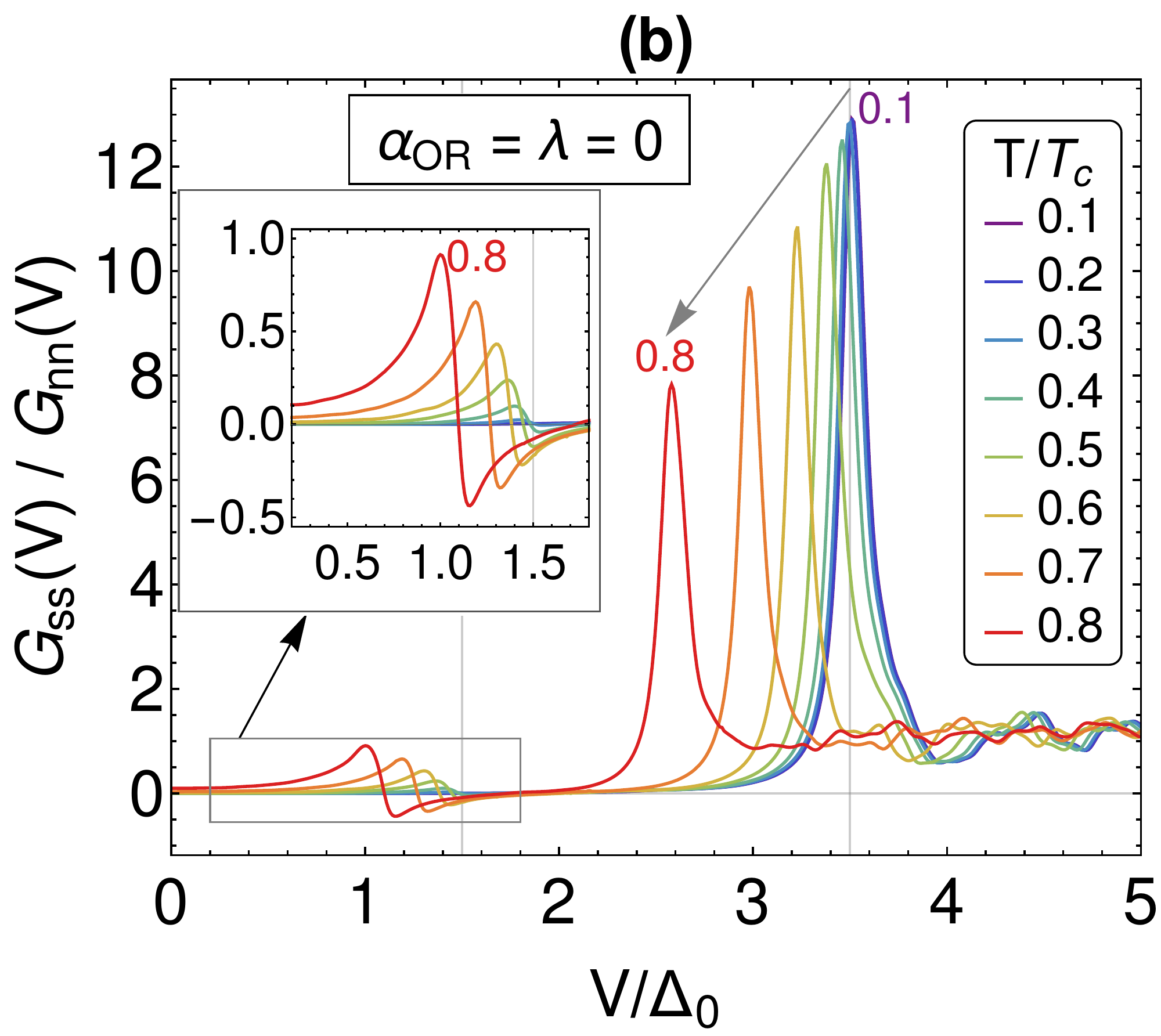}
\protect\caption{Normalized tunneling current $I_{ss}(V)$ (a) and  conductance $G_{ss}(V)$ (b) for an SIS$_2$ junction in the absence of an applied electric field (i.e. for $\alpha_{\text{OR}}=\lambda=0$) as a function of temperature. For S$_2$ we are considering a conventional BCS superconductor with $\Delta_{S_2}=2.5 \Delta_0$. The vertical gray lines mark the position of the bias voltage $V_{-}=\Delta_{S_2}-\Delta_S$ and $V_{+}=\Delta_S+\Delta_{S_2}$ at zero temperature, with $\Delta_S=\Delta_0$. In the insets we show a zoom in the voltage bias region where the matching peaks occur.}
\end{figure}
%%%%%

To proceed further, we consider representative profiles of the superconducting order parameters for the various bands including the states with $\pi$-pairing (Fig. 1 (c)-(e)) and correspondingly determine the layer dependent density of states by computing the energy spectra and the eigenvectors of the Hamiltonian. 

Tunneling spectroscopy in junctions or in scanning tunneling experiments is an important probe of the density of states (DOS) of target materials. The current across the junction at a finite applied bias $V$ is generally expressed as a convolution of the DOS of the normal metal electrode, $N_{n}(E)$, that one of the material upon examination, $N_{s}(E)$, and the Fermi-Dirac function $f(E)$,
\begin{eqnarray}
I(V)=\int N_{n}(E) N_{s}(E+e V) [f(E)-f(E+e V)] dE
\end{eqnarray}
\noindent while the conductance is given by $G(V)=\frac{d I(V)}{dV}$. For our purposes we assume that the normal metal in the SIN has a constant density of states in the range of investigated voltages while the non-gated superconducting electrode in the SIS is described by a conventional BCS-type density of states. 
  
The presence of the Fermi-Dirac distribution in the expression of the SIN tunneling conductance limits the energy resolution (i.e. $\sim k_{\mathrm{B}} T$), with the thermal smearing that often precludes the detection of features with low intensity and small energy separation in the density of states. Indeed, the inspection of the DOS for the superconducting electrode shows that there are more fringes and spectral features at energies that are smaller than the thermal smearing. 
In Fig. \ref{f2} we report the thermal evolution of the SIN tunneling conductance by assuming that the superconducting electrode is described by the tight-binding model Hamiltonian of Eq. (2). Here, we take a representative case with $n_z=6$, the analysis with a larger number of layers does not affect the overall spectroscopic outcome. Additionally, we assume that the superconducting order parameter is spatially uniform and independent of the orbital index. This position has two main motivations. It is substantially consistent with the typical profiles that are obtained by fully minimizing the free energy with layer dependent superconducting order parameters, as explicitly depicted in Fig. \ref{f1}(c)-(e). 
Additionally, in order to single out the spectroscopic fingerprints that uniquely arise from the modification of the electric field, through the orbital Rashba interactions, it is convenient to neglect the small amplitude differences between the superconducting order parameters in different bands. This strategy is quite neat because it helps in clarifying the impact of the electric field without mixing with other effects related to pairing amplitude variation. Indeed, the presence of multi-gap structure of superconductors leads to other features in the tunneling conductance and its spectroscopy detection is an intricate problem that stands alone even without the application of the electric field \cite{Noat2010}. 

Furthermore, for the junction's electrode subjected to the electrostatic gating, we explicitly compute the density of states for a multilayered superconductor as described by the model Hamiltonian in Eq. 1. 
Here, the analysis of the electronic energy states and the corresponding eigenfunctions is performed within the whole Brillouin zone and, then, by evaluating the momentum integration of the Bogoliubov energies taking into account the band and layer dependence of the superconducting order parameters (Fig. 1 (c)-(e)).

The outcome in Fig. \ref{f2} sets the reference for our analysis for the configuration with vanishing electric field. As expected the conductance has a peak at voltages $V_{\mathrm{max}}\sim \Delta_0$ and the position of $V_{\mathrm{max}}$ moves to values of $\sim 1.5 \Delta_0$ while approaching $T_{\mathrm{c}}$. Furthermore, there is a full gap at low temperature that fills up uniformly from zero to one as the temperature reaches the critical temperature.

Replacing the normal electrode with a superconducting one in the SIS geometry, one can overcome the thermal broadening issues by exploiting the non-linearities of the density of states of the probing superconducting electrode. Indeed, it is well known that the SIS spectroscopy can lead to a significant enhancement of the resolution of the tunneling spectroscopy. To set the reference we consider an SIS heterostructure where one side of the junction, labeled as $S_2$, is described by a conventional BCS-type density of states including the Dynes phenomenological parameter $\Gamma$, $N_{S_2}(E)=N_0 \mathrm{Re}[\frac{E+i \Gamma}{\sqrt{(E+i \Gamma)^2-\Delta_2^2}}]$. 
The parameter $\Gamma$ is usually employed to quantify the consequences of the pair breaking processes, while $N_0$ is the value of the density of states at the Fermi level in the normal metal phase.

In Fig. 3 we show the SIS tunnel conductance for the case of zero applied gating (i.e. $\alpha_{OR}=\lambda=0$). The outcome describes the typical profile of the SIS tunnel current and conductance. Here, we assume that the gap amplitude of the electrically gated superconductor (S) is $\Delta_S=\Delta_0$ while that one of the conventional electrode (S$_2$) is $\Delta_{S_2}=2.5 \Delta_0$. The current profile exhibits the typical peak when the applied voltage matches the difference of the gaps, $V_{-}=\Delta_{S_2}-\Delta_S$ ($1.5 \Delta_0$), and the rapid increase at the sum of the gaps, $V_{+}=\Delta_S+\Delta_{S_2}$ ($3.5 \Delta_0$). The peak at $V_{-}$ becomes visible and more pronounced with reduced temperatures $T/T_c$ getting above about 0.5. This structure in the current profile gives a non-monotonous behavior for the conductance with a change of sign close to $V_{-}$. Furthermore, the rapid upturn of the current corresponds to a peak in the conductance that moves to lower voltages while increasing the temperature upon reaching $T_c$. We notice that for voltages above $V_{+}$ there are small amplitude oscillations in the conductance that reflect the intrinsic features of the electronic structure in the layered superconducting thin film described by the Hamiltonian in Eq. 1.   

%%%%
\begin{figure*}[bt]
\includegraphics[width=0.3\textwidth]{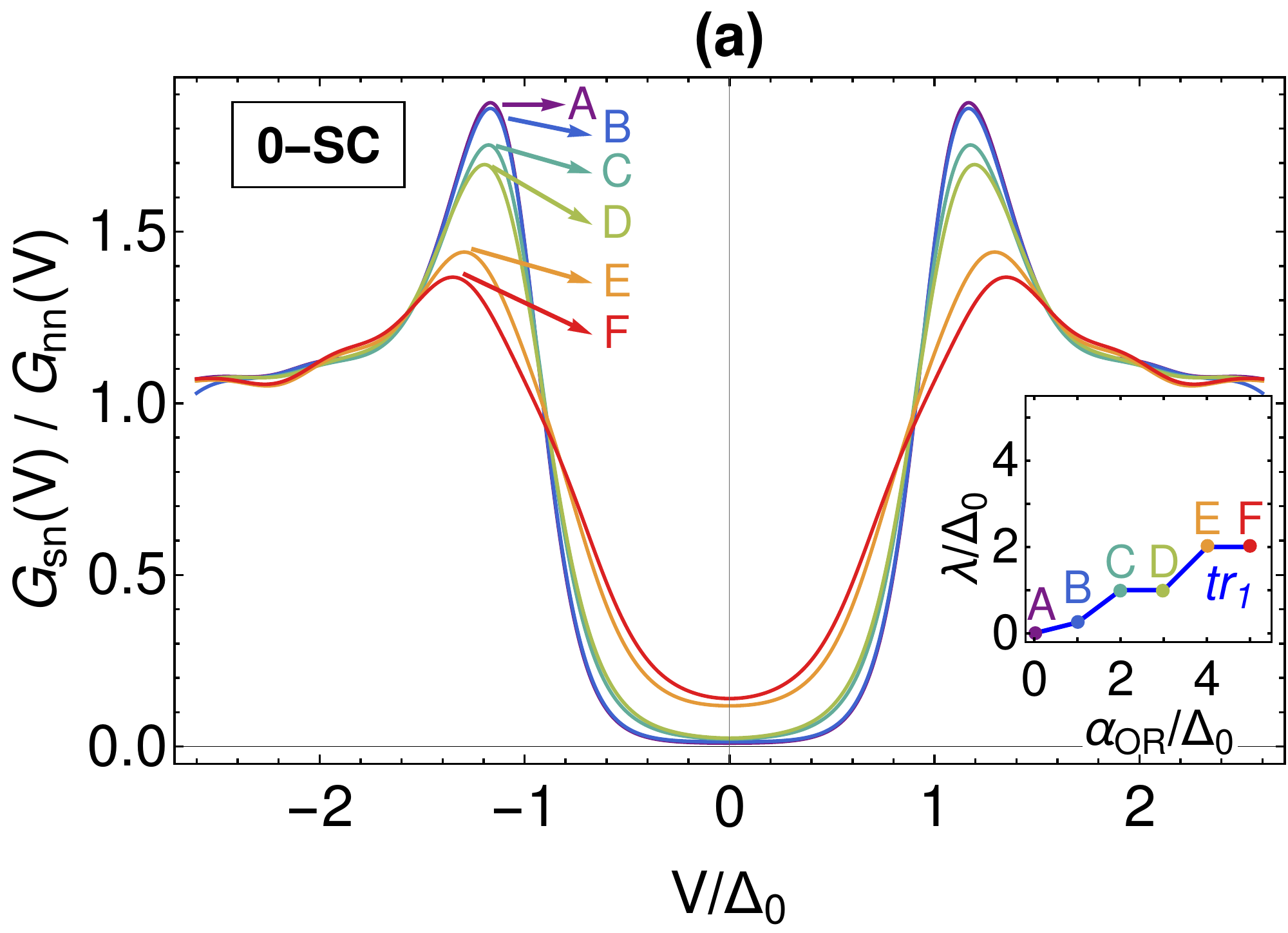} \hspace{0.4cm}
\includegraphics[width=0.3\textwidth]{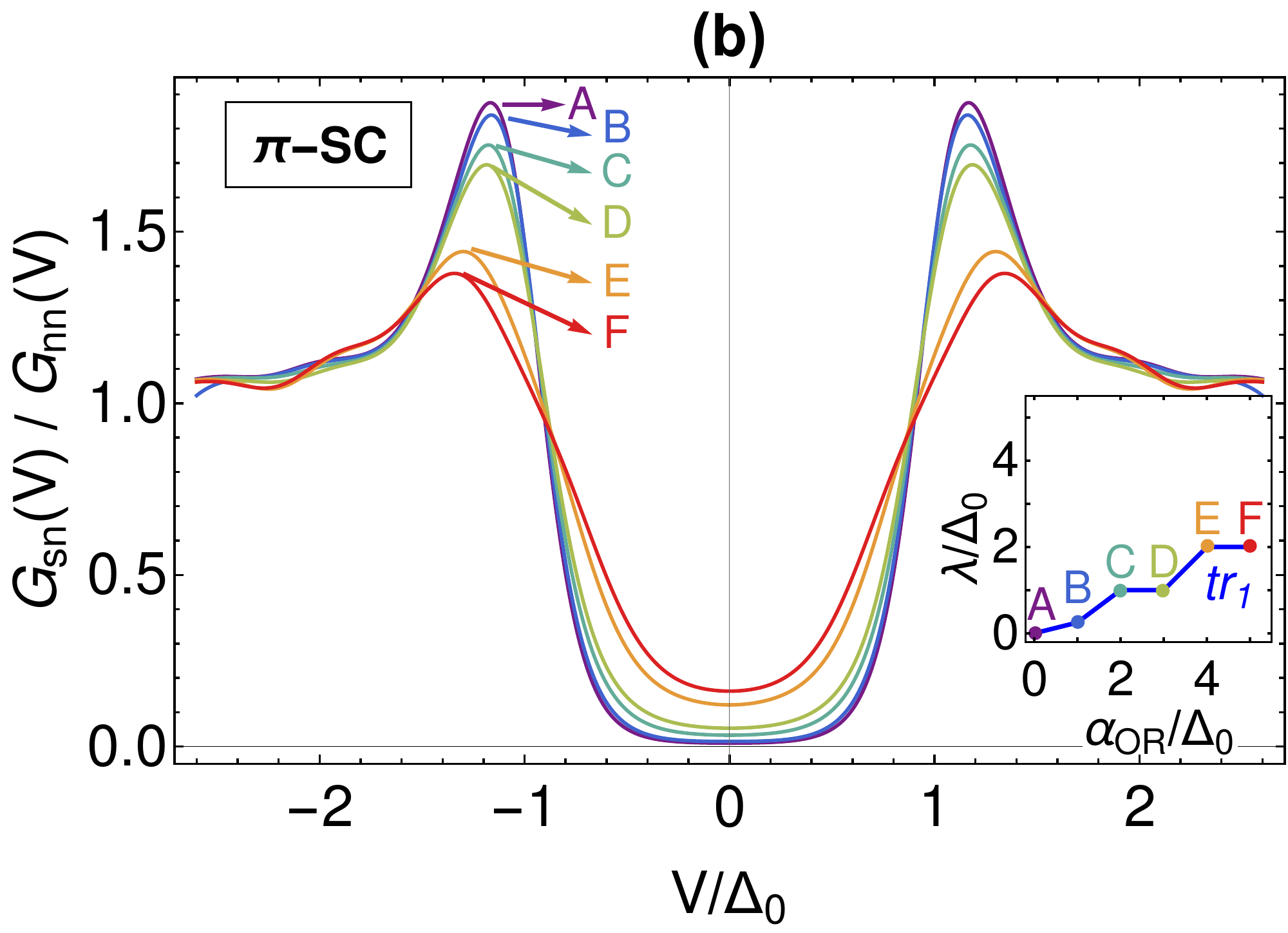} \hspace{0.4cm}
\includegraphics[width=0.3\textwidth]{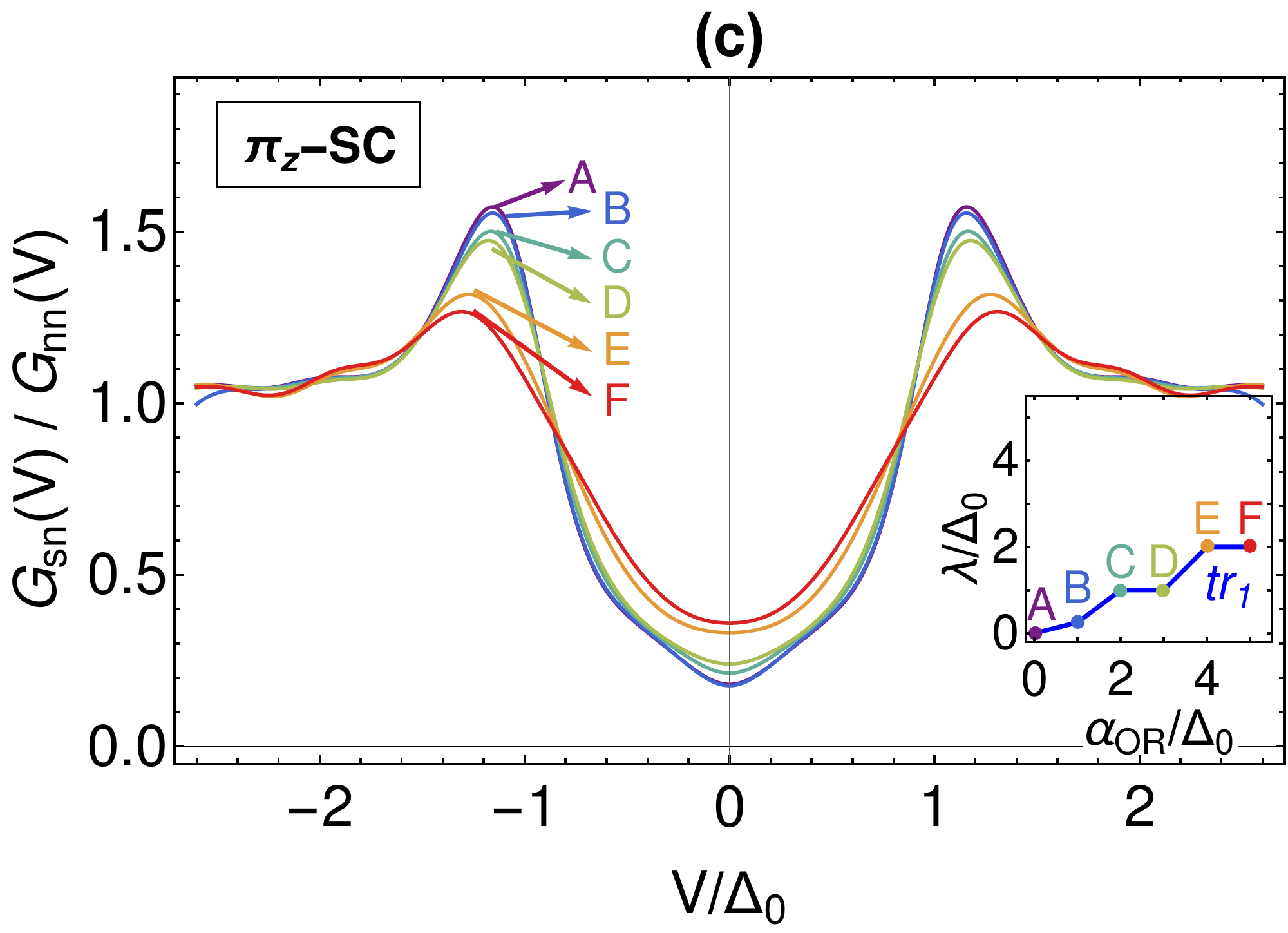} \\
\vspace{0.5cm}
\includegraphics[width=0.3\textwidth]{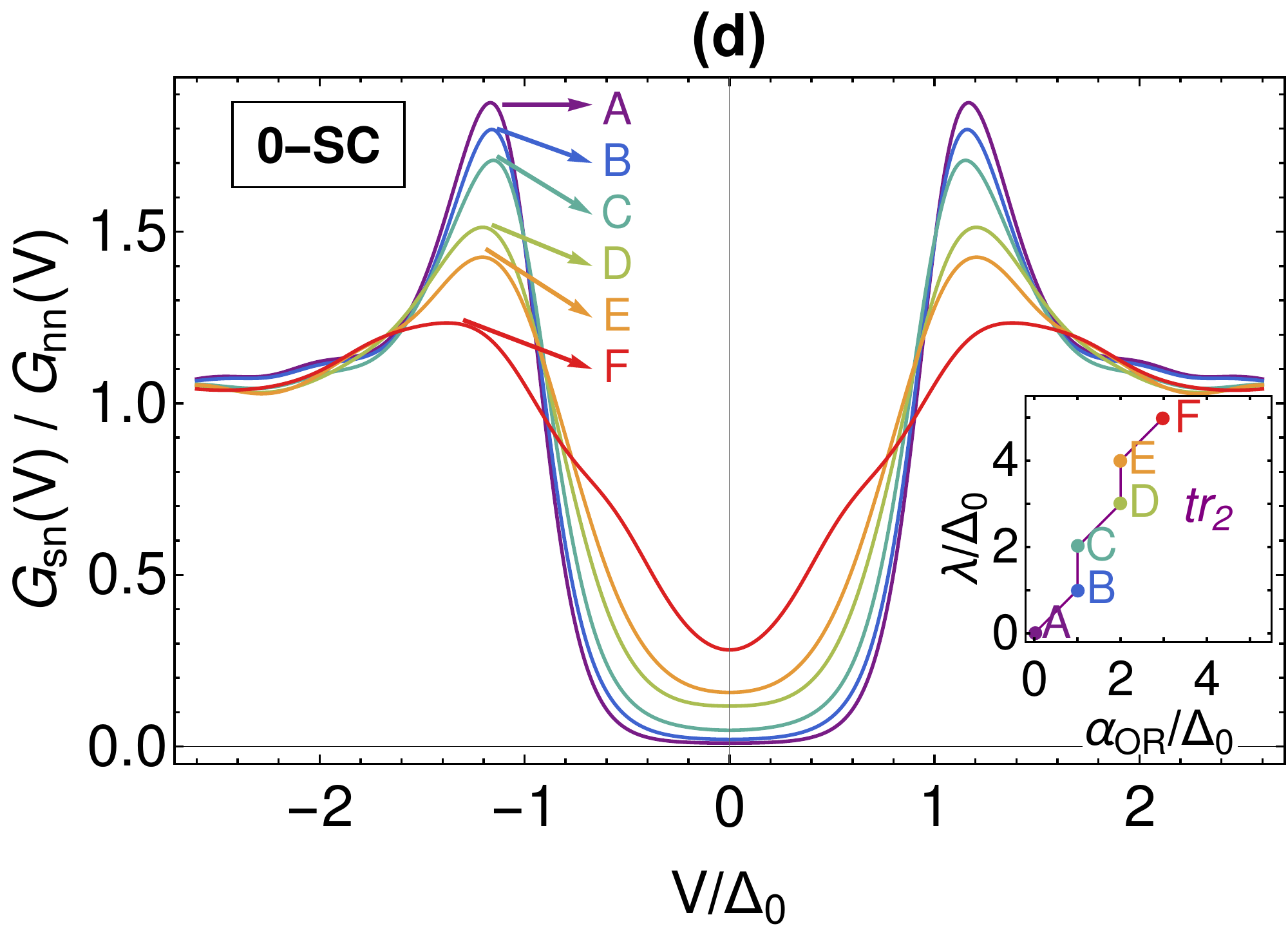} \hspace{0.4cm}
\includegraphics[width=0.3\textwidth]{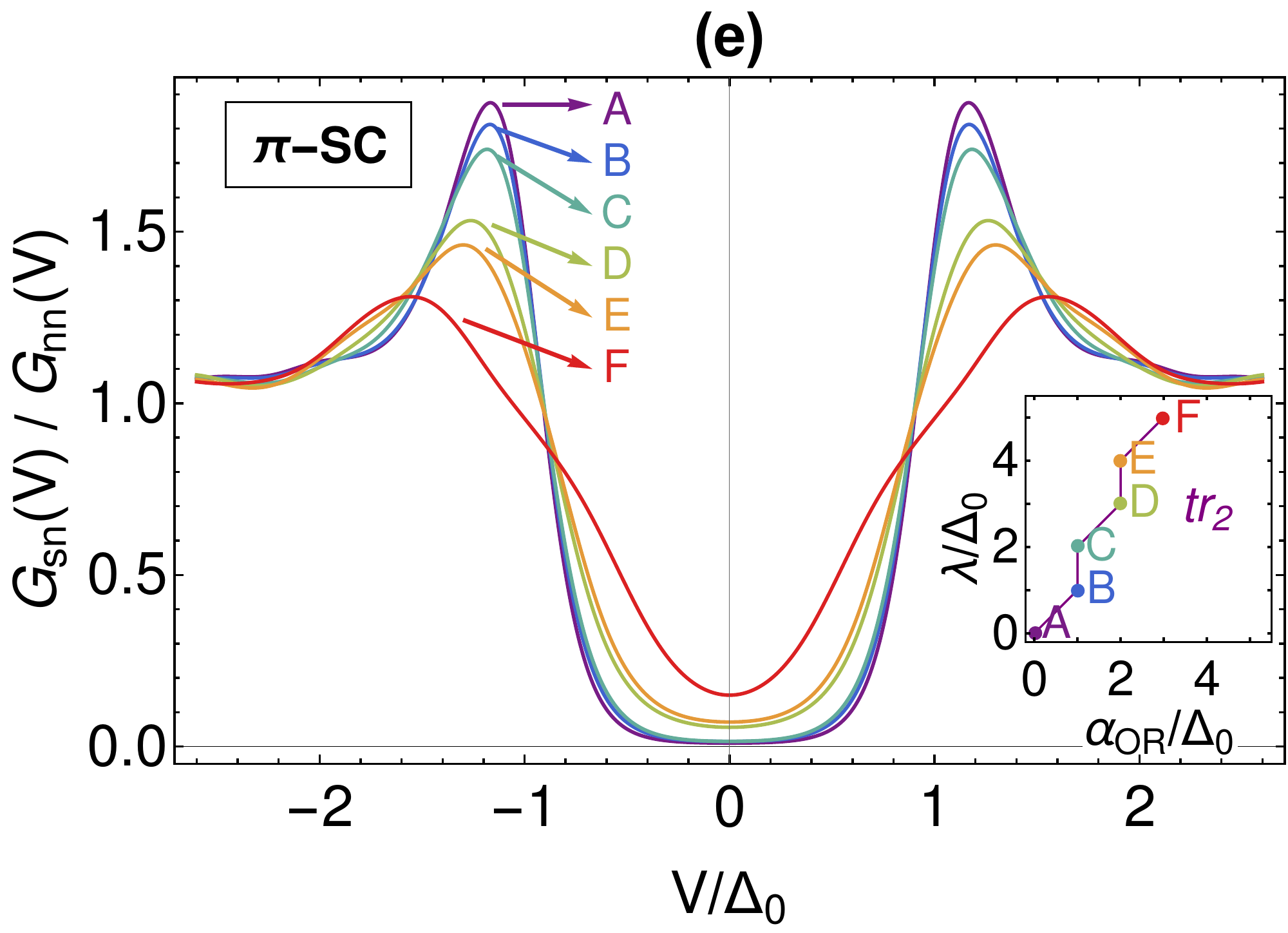} \hspace{0.4cm}
\includegraphics[width=0.3\textwidth]{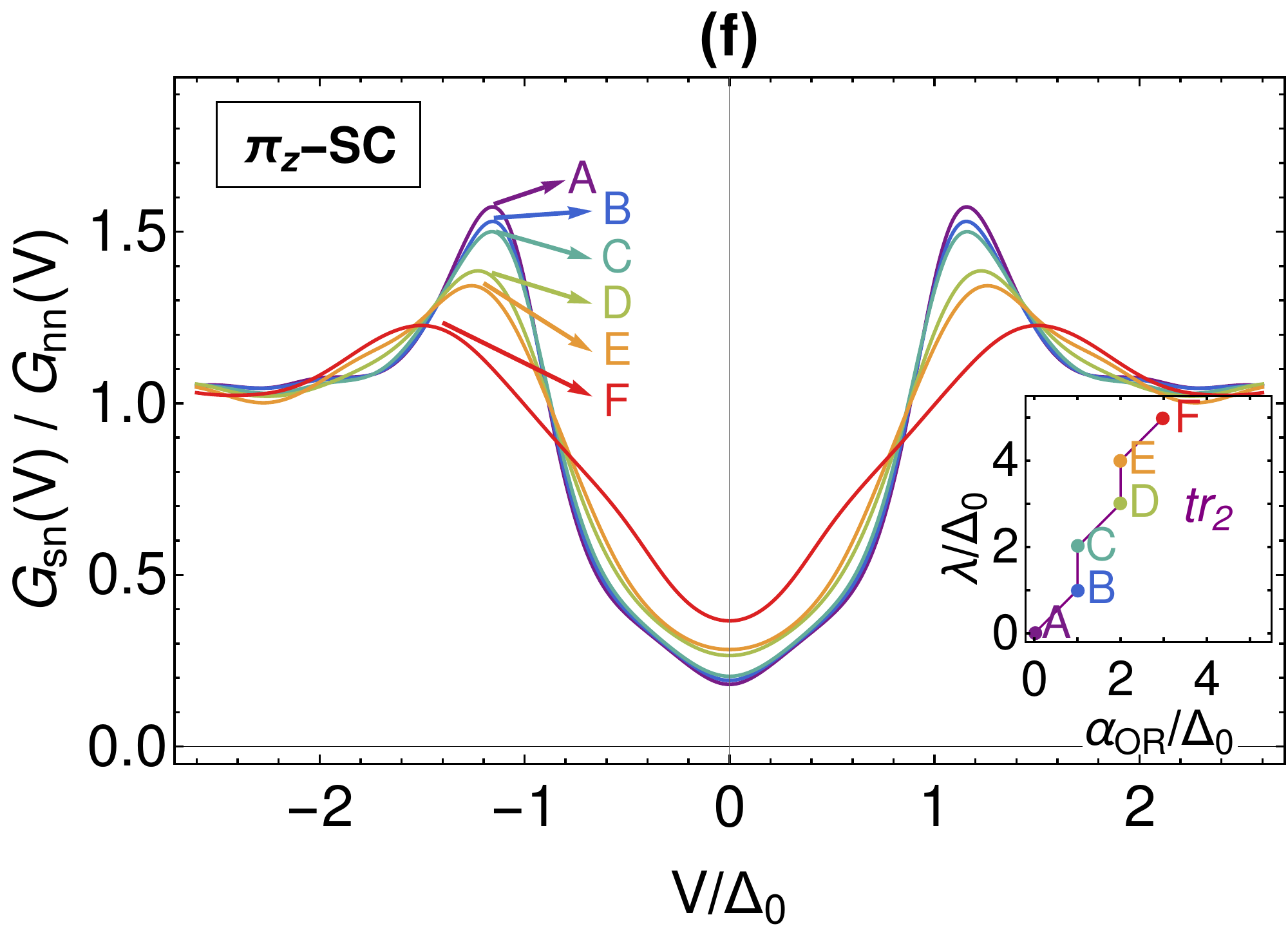}
\protect\caption{Tunneling conductance $G_{sn}(V)$, normalized to the normal one $G_{nn}(V)$, for an SIN junction  at low temperature ($t=T/T_c=0.2$),  for the 0-SC, $\pi$-SC  and $\pi_{z}$-SC phases along two given representative trajectories in the parameters space corresponding to the values of the intra- ($\alpha_{OR}$) and inter-layer ($\lambda$) orbital Rashba coupling that have been employed for the computation: (a-c)  trajectory with $\lambda<\alpha_{{\text{OR}}}$; (d-f)  path with $\lambda>\alpha_{{\text{OR}}}$. The gated superconductor has a layered geometry with $n_z=6$ layers.}
\label{SIN}
\end{figure*}
%%%%

\section{Results}

In this Section we present the evolution of the SIN and SIS tunneling conductance due to the application of an electrostatic gating onto one superconducting electrode constituting the junction. The gating can drive the superconducting state, resulting into 0-, $\pi$-, or $\pi_z$- configurations with orbitally polarized surface states (Fig. \ref{f1}). Since the amplitude of the microscopic parameters directly affected by the electric field is generally proportional to $E_s$ with a factor that depends on materials characteristics, we have taken different trajectories in the phase space spanned by $\alpha_{\text{OR}}$ and $\lambda$. The focus is on the low temperature regime because it can allow to single-out the consequences of the electrostatic gating and distinguish from the canonical thermal effects in conventional junctions.

\subsection{SIN conductance}

Let us start by considering the SIN tunneling conductance for two representative trajectories in the phase space assuming that a variation of $E_s$ allows us to move along a path with $\lambda$ that is smaller (Fig. \ref{SIN}(a)-(c)) or greater (Fig. \ref{SIN}(d)-(f)) than $\alpha_{\text{OR}}$, respectively. The range of variation for $\alpha_{\text{OR}}$ and $\lambda$ is taken from 0 to 5 $\Delta_0$, with $\Delta_0$ setting the superconducting gap in absence of external perturbations. 
This is an energy range for the electrostatic gating that is suitable for the experimental observations and also microscopically consistent for covering the allowed phase transitions from 0- to $\pi$-paired states. We remind that $\lambda$ and $\alpha_{\text{OR}}$ depend on $E_s$ in such a way that larger amplitudes for $E_s$ would typically lead to a normal phase with significant orbital polarization at the Fermi level. 
The strategy we follow is to explore the response of the three different phases (i.e. 0-, $\pi$-, and $\pi_z$-SC) that can be achieved in a multi-band superconductor upon the application of an electrostatic gating independently of their stability in the phase diagram. This general approach allows us to get a complete view of the behavior of the tunneling conductance for each superconducting state in the whole phase space. 
Beginning from the 0-SC state we find that the application of the electric field generally leads to a filling up of the gap (Fig. \ref{SIN}(a),(d)). Depending on the ratio between $\lambda$ and $\alpha_{\text{OR}}$ there can be a high ($\lambda>\alpha_{\text{OR}}$) or low ($\lambda<\alpha_{\text{OR}}$) rate of increase of the quasi-particle in-gap population. This result clearly indicates that the surface inter-layer coupling $\lambda$ is a source of depairing by orbitally polarizing the electronic configurations at the Fermi level. This is expected indeed because the $\lambda$ term acts as an effective orbital current and can locally break the time reversal symmetry. 
Here, we notice that in the regime of large $\lambda$ (Fig. \ref{SIN}(d)) the filling of the gap is accompanied by the formation of a bump at voltages that are below the gap edge of the order of $\sim 0.5 \Delta_0$. Since the impact of the electrically driven couplings is primary on the surface, we argue that such feature mostly originates from significant surface reconstruction of the in-gap electronic states.  

Another remarkable spectroscopic aspect of the electrically driven SIN junction refers to the evolution of the maximum of conductance. Starting from the case at zero applied electric field we observe that the peak tends to move to voltages larger than $\Delta_0$ while approaching the boundaries of the explored phase space. This shift is generally accompanied by a suppression of the amplitude of the conductance peak. A close inspection of the evolution indicates a non-monotonous behavior of the voltage position ($V_{{\text{max}}}$) of the conductance peak. In fact, in the regime of $\lambda$ and $\alpha_{{\text{OR}}}$ being comparable to $\Delta_{0}$, corresponding to the $A$-$C$ path, with $\lambda> \alpha_{{\text{OR}}}$ in Fig.\ref{SIN}(d), $V_{\text{max}}$ tends to decrease to lower voltage intensity. The overall trend keeps being non-monotonous also along the $D$-$E$ path. Although the inward shift is typically small, this is a qualitative distinctive mark that can be also experimentally detectable.

Let us now move to the $\pi$-paired phase. In this context, we deal with two possible configurations with uniform ($\pi$-SC) or spatially inhomogeneous ($\pi_z$-SC) order parameters along the $z$-direction of the superconducting thin film (Fig. \ref{f1} (d),(e)). 
For the uniform $\pi$-SC the tunneling conductance exhibits similar features of those found for the 0-SC case. Indeed, the increase of the electric field amplitude along the two identified trajectories in the $(\lambda,\alpha_{{\text{OR}}})$ phase space generally yields a growing in-gap conductance contributions (Fig. \ref{SIN}(b),(e)). Furthermore, the in-gap bump obtained approaching the point $F$ in the $\lambda> \alpha_{{\text{OR}}}$ path (Fig. \ref{SIN}(e)) is robust as it occurs in the $\pi$-SC state too. When considering the evolution of the maximum of conductance, some differences are encountered with respect to the 0-SC configuration. Indeed, we find that the peak moves monotonously to high voltages irrespective of the type of trajectory in the phase space. In particular, we find that for the path with $\lambda> \alpha_{{\text{OR}}}$ the maximum can shift up to $V\sim 1.5 \Delta_0$. Thus, in the uniform $\pi$-SC phase the conductance peak is suppressed and can move to large voltage amplitudes. The shift is particular pronounced for the trajectory with $\lambda> \alpha_{{\text{OR}}}$. 

Finally, we study the SIN conductance for a superconducting electrode that is marked by $\pi_z$-pairing. In this configuration, we have that 0- and $\pi$-SC coexist in the thin film as they occur in different layers along the confining direction. More specifically, for the analyzed layered superconductor the top/bottom surface layers have no interband phase shift, while the remaining layers are marked by $\pi$-pairing. Hence, apart for the intrinsic inter-band phase shift at the Fermi level, the superconducting order parameter exhibits an amplitude spatial modulation with a sign change when moving along the $z$-direction. This is similar to unconventional pairing with spatial modulation of the amplitude as for the case of the Larkin-Ovchinnikov state \cite{Larkin1965}, such variation is expected to result into nodal excitations or to generally bring an effective suppression of the superconducting gap.
This is indeed obtained when considering the limit of vanishing electric field. The zero bias conductance in that case is large indicating a substantial contribution of in-gap electronic states nearby the Fermi level. Then, a variation of the orbital Rashba couplings drives an increase of the in-gap conductance which however is of the same order of magnitude of that observed in the uniform 0- and $\pi$-SC phases. 
The evolution of the peak in the conductance also shows a trend with an increase of the voltage and a suppression of the amplitude as one moves along the two selected trajectories. 
In particular, the shift of the conductance maximum is amplified when proceeding along the path corresponding to $\lambda> \alpha_{{\text{OR}}}$ reaching a value of $V_{\text{max}} \sim 1.5 \Delta_0$. 

At this point it is instructive to compare the profile of the conductance as a function of the applied electrostatic gating (Fig. \ref{SIN}) with that one obtained by varying the temperature (Fig. \ref{f2}). In Fig. \ref{f2} we have shown the canonical temperature dependence of an SIN junction for a conventional BCS superconductor.
The thermal effects induce a filling up of the in-gap conductance with a suppression of the maximum amplitude and a shift of its peak position at high voltages while approaching the superconducting transition temperature $T_c$. We notice that the increase of the conductance is quite uniform within the gap while the zero bias amplitude approaches its normal value close to $T_c$.
On the other hand, it is unexpected that at given temperature the application of an electrostatic gating leads to a similar qualitative trend when compared to the thermal drive. This implies that the SIN tunneling conductance with the superconducting electrode subjected to an external electric field can yield outcomes which in principle are difficult to get distinguished from thermal effects.  
In particular, the analogies are evident for the $\pi$-phases. Instead, the non-monotonous evolution of the maximum of the conductance upon the application of an electrostatic gating for the 0-SC represents a mark of unique electric effects in SIN tunneling detection, although such consequence is expected only in the regime of small electric field amplitudes.

%%%%
\begin{figure*}[bt]
\includegraphics[width=0.3\textwidth]{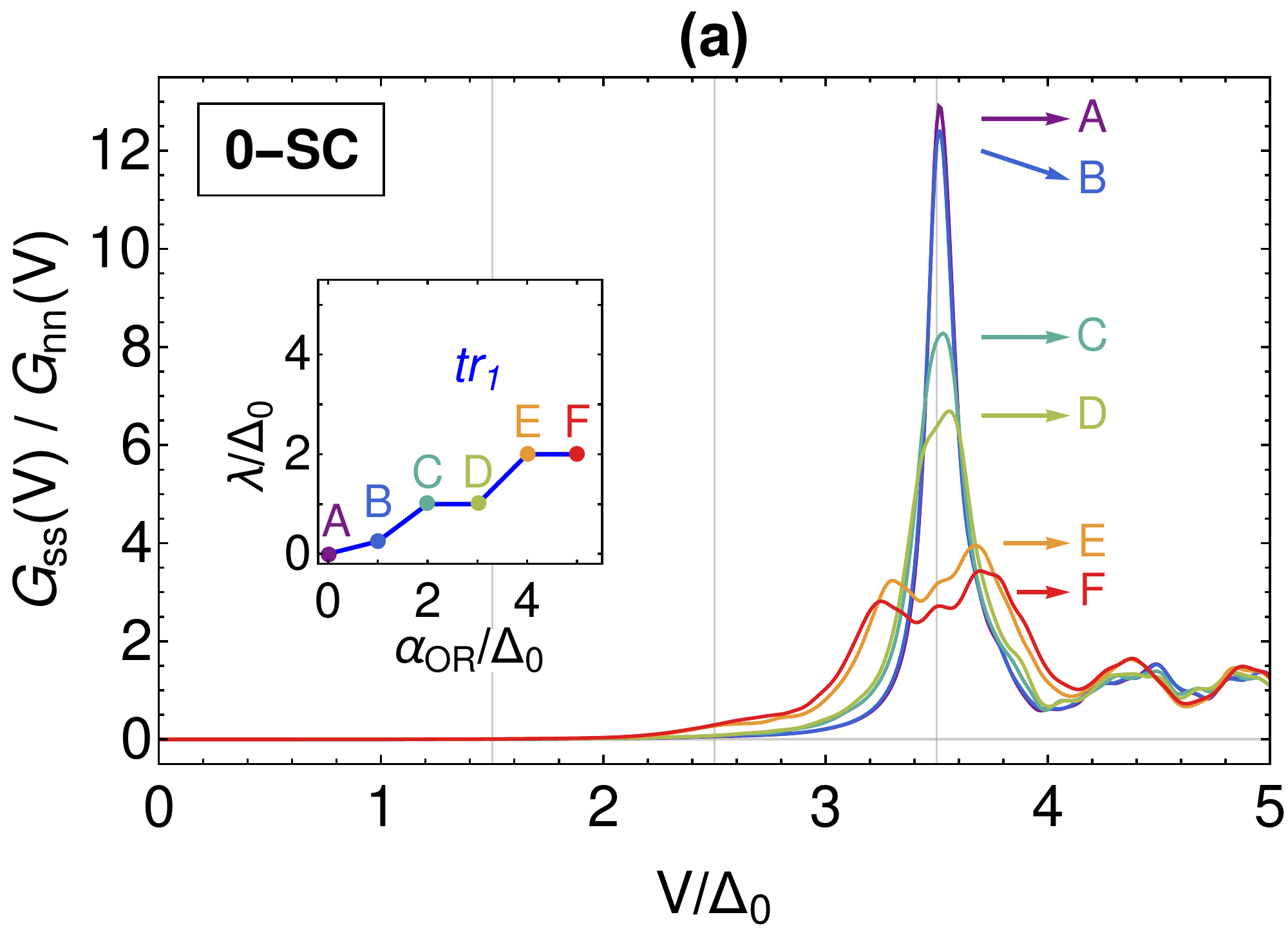} \hspace{0.4cm}
\includegraphics[width=0.3\textwidth]{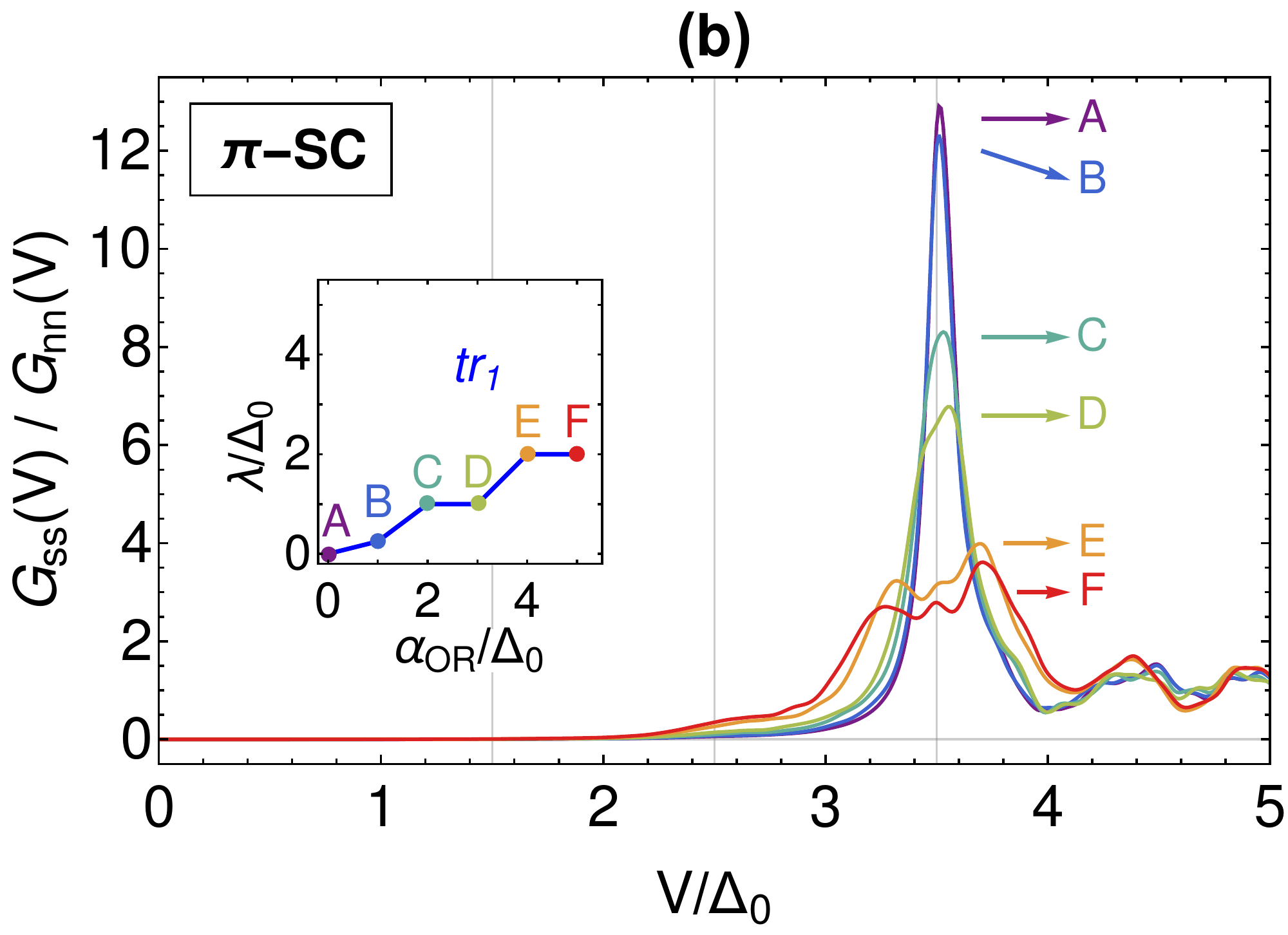} \hspace{0.4cm}
\includegraphics[width=0.3\textwidth]{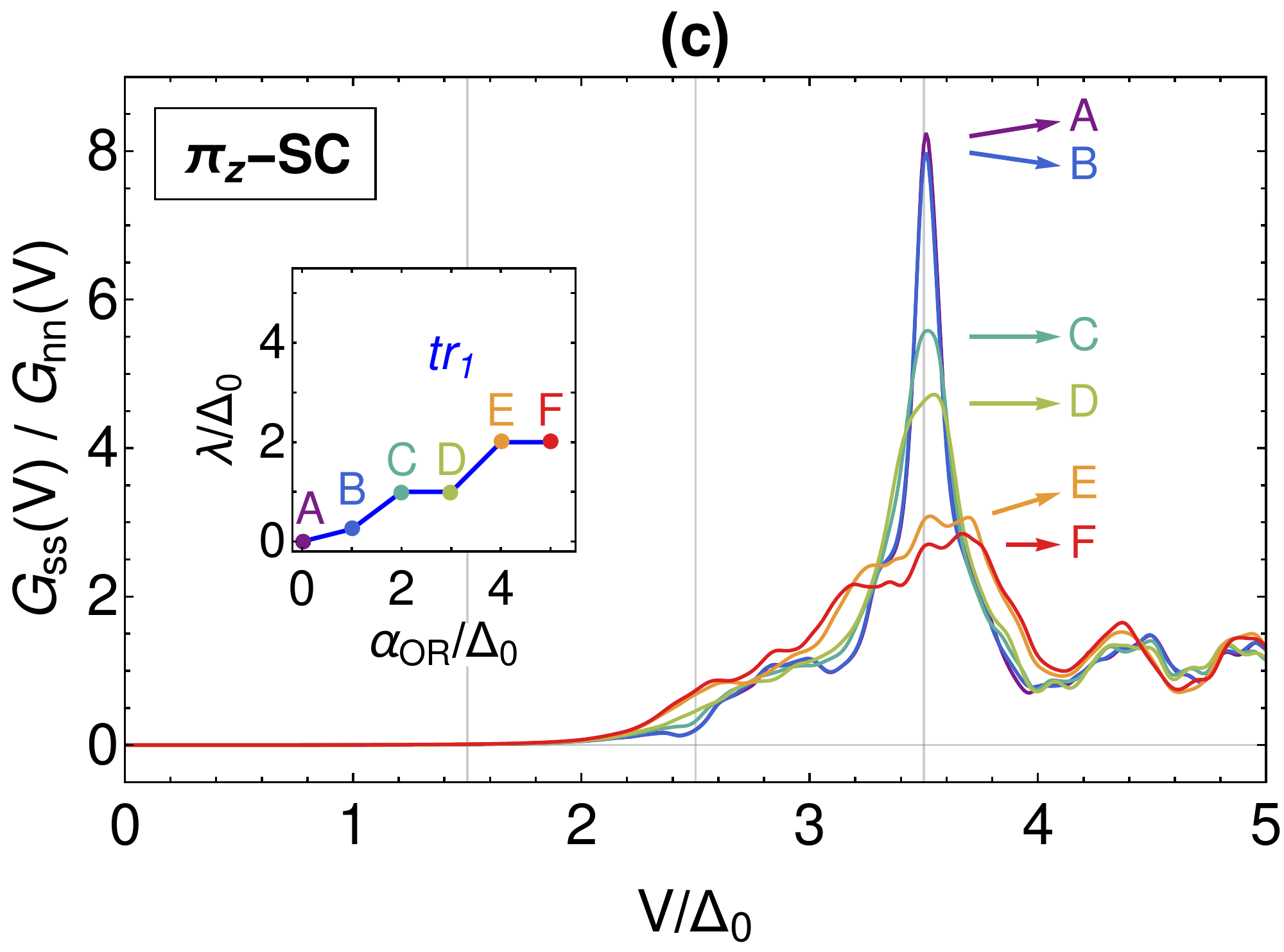} \\
\vspace{0.5cm}
\includegraphics[width=0.3\textwidth]{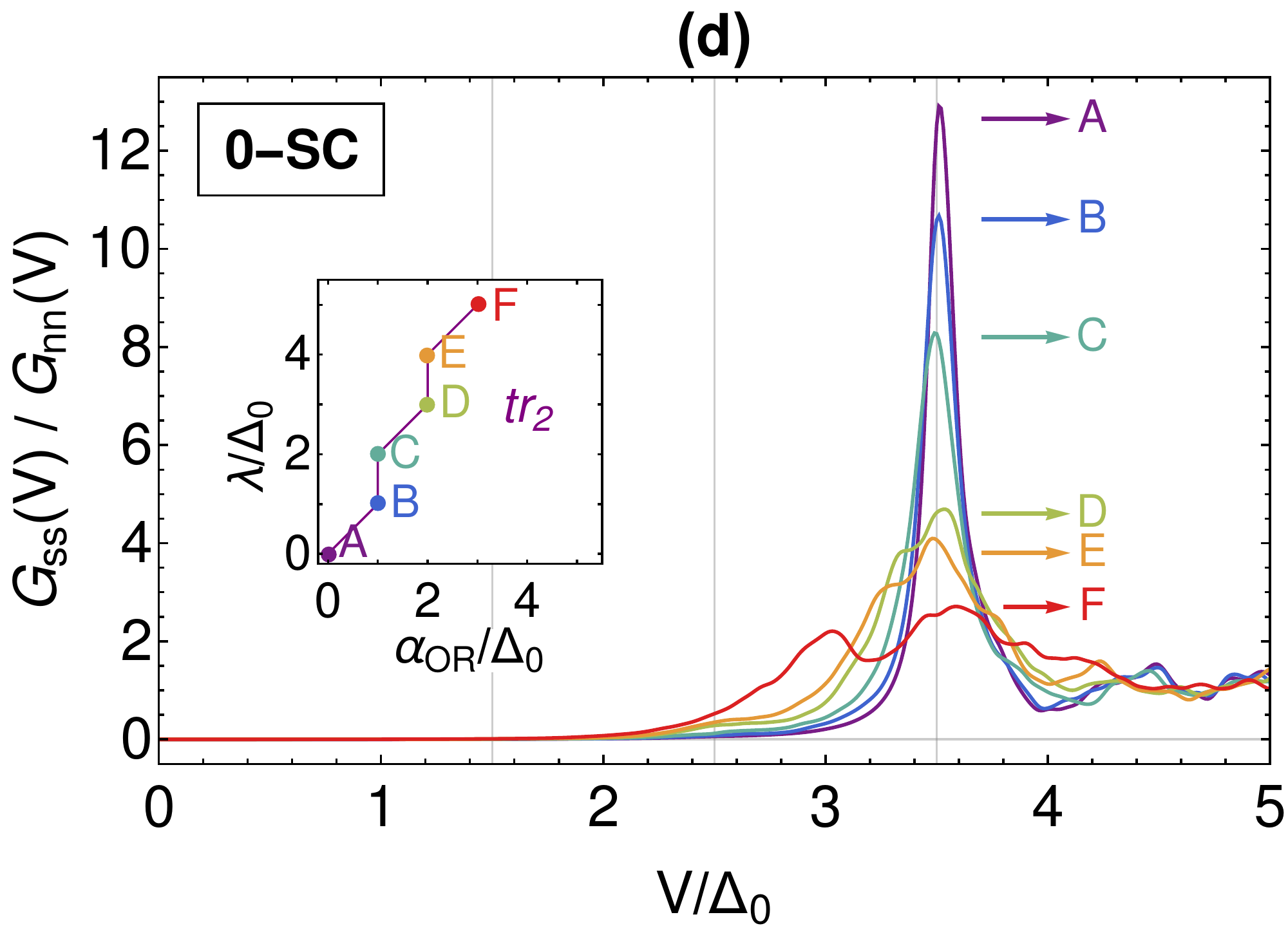} \hspace{0.4cm}
\includegraphics[width=0.3\textwidth]{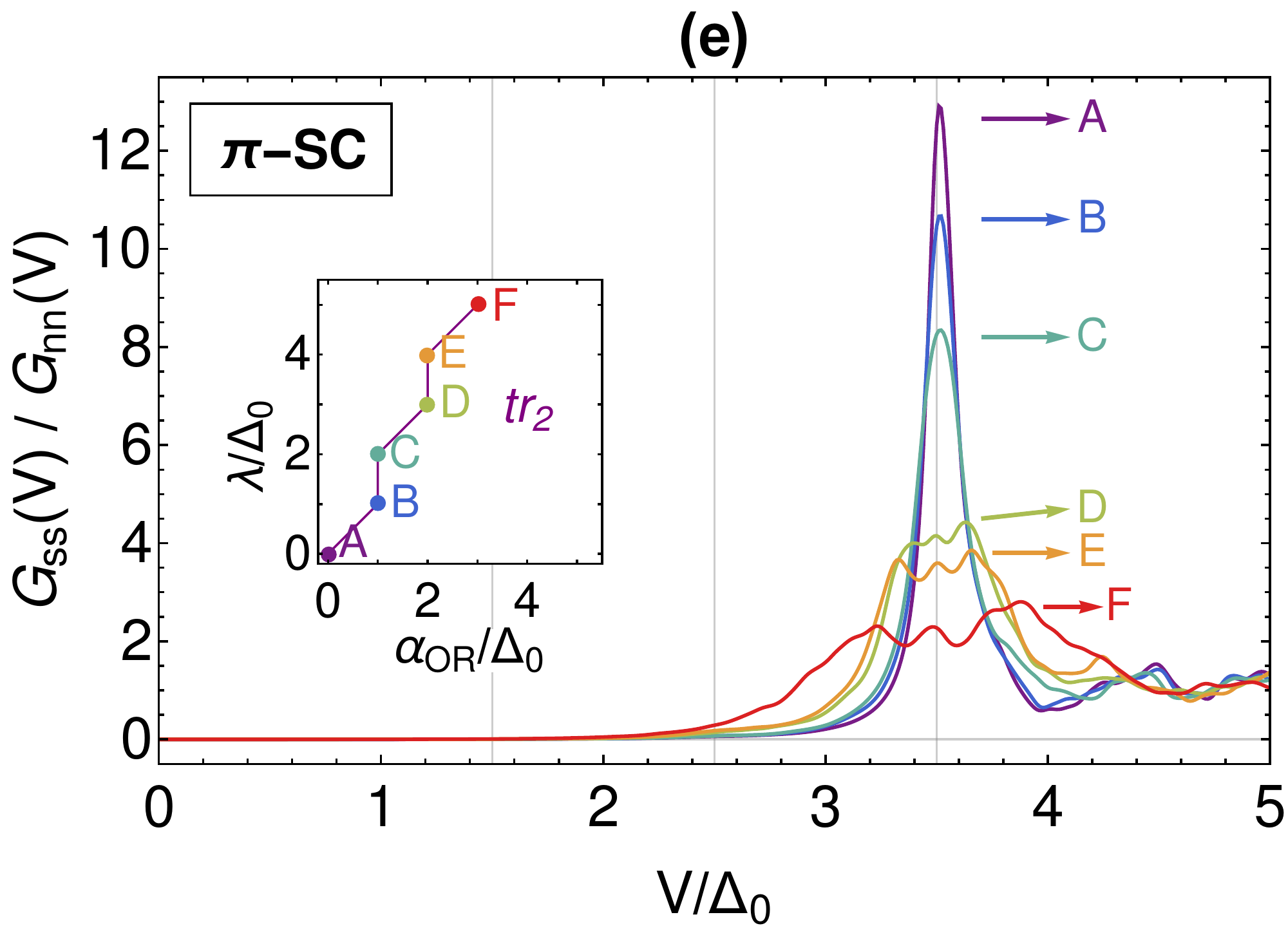} \hspace{0.4cm}
\includegraphics[width=0.3\textwidth]{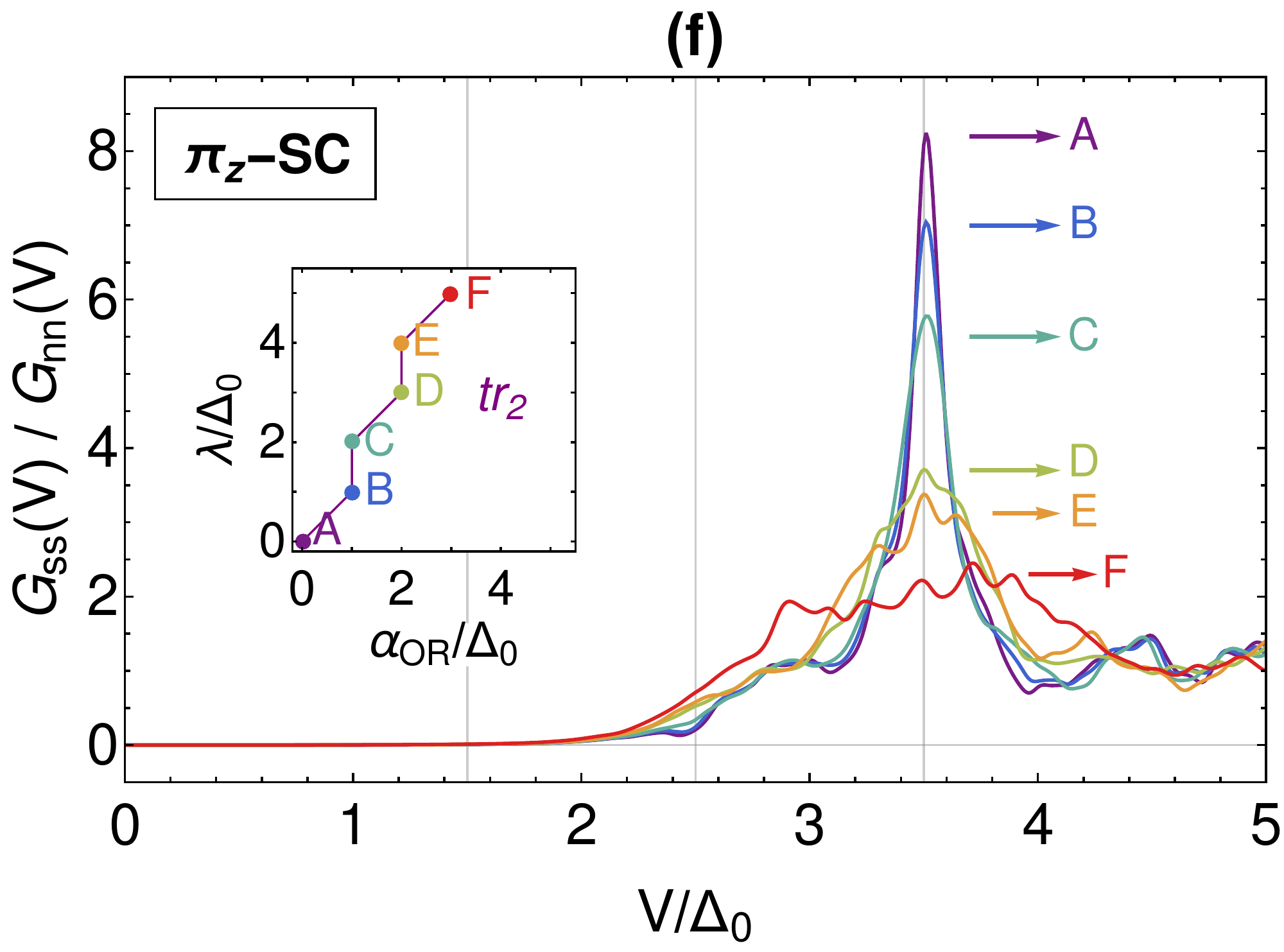}
\protect\caption{Superconducting tunneling conductance $G_{ss}(V)$, normalized to the normal state configuration, for an SIS$_2$ junction  at low temperature ($t=T/T_c=0.2$), for 0-SC, $\pi$-SC  and $\pi_{z}$-SC phases along two given representative trajectories in the parameters space corresponding to the values of the intra- ($\alpha_{{\text{OR}}}$) and inter-layer ($\lambda$) orbital Rashba coupling that have been used for the computation: (a-c)  path with $\lambda<\alpha_{OR}$; (d-f)  trajectory with $\lambda>\alpha_{{\text{OR}}}$. For S$_2$ we are considering a BCS-type superconductor with $\Delta_{S_2}=2.5 \Delta_0$. The vertical gray lines mark the position of $V_{-}=\Delta_{S_2}-\Delta_S$, $V_{2}=\Delta_{S_2}$ and $V_{+}=\Delta_S+\Delta_{S_2}$, with $\Delta_{S}=\Delta_0$. The gated superconductor S has a layered geometry with $n_z=6$ layers.}
\label{SIS}
\end{figure*}
%%%%

\subsection{SIS conductance}

Having shown that the SIN tunneling conductance cannot directly provide distinctive features to disentangle the role of electrostatic gating from thermal effects, we discuss the major differences that can be extracted once considering the SIS spectroscopic response.

As for the SIN tunneling, the strategy is to search for clearcut fingerprints which might be used to assess the nature of electrically driven superconducting phases by exploring the 0-, $\pi$- and $\pi_z$-SC states. 
In Sect. II we have shown that for SIS$_2$ heterostructures with inequivalent superconducting electrodes, a maximum in the conductance is expected to occur at voltage positions that correspond to the difference and sum of the superconducting gaps, with the former being pronounced only above an effective temperature. Hence, it is relevant to ask about the impact of the electric field in the energy range where the matching peak and the characteristic conductance features manifest. Furthermore, taking a temperature where the lowest voltage matching peak is vanishing and not detectable, it is also worth evaluating whether the electric field can induce it, thus mimicking thermal effects as in SIN tunneling spectroscopy, or bring substantially different consequences.

%%%%%%%%%%%%%
\begin{figure*}
\includegraphics[width=0.16\textwidth]{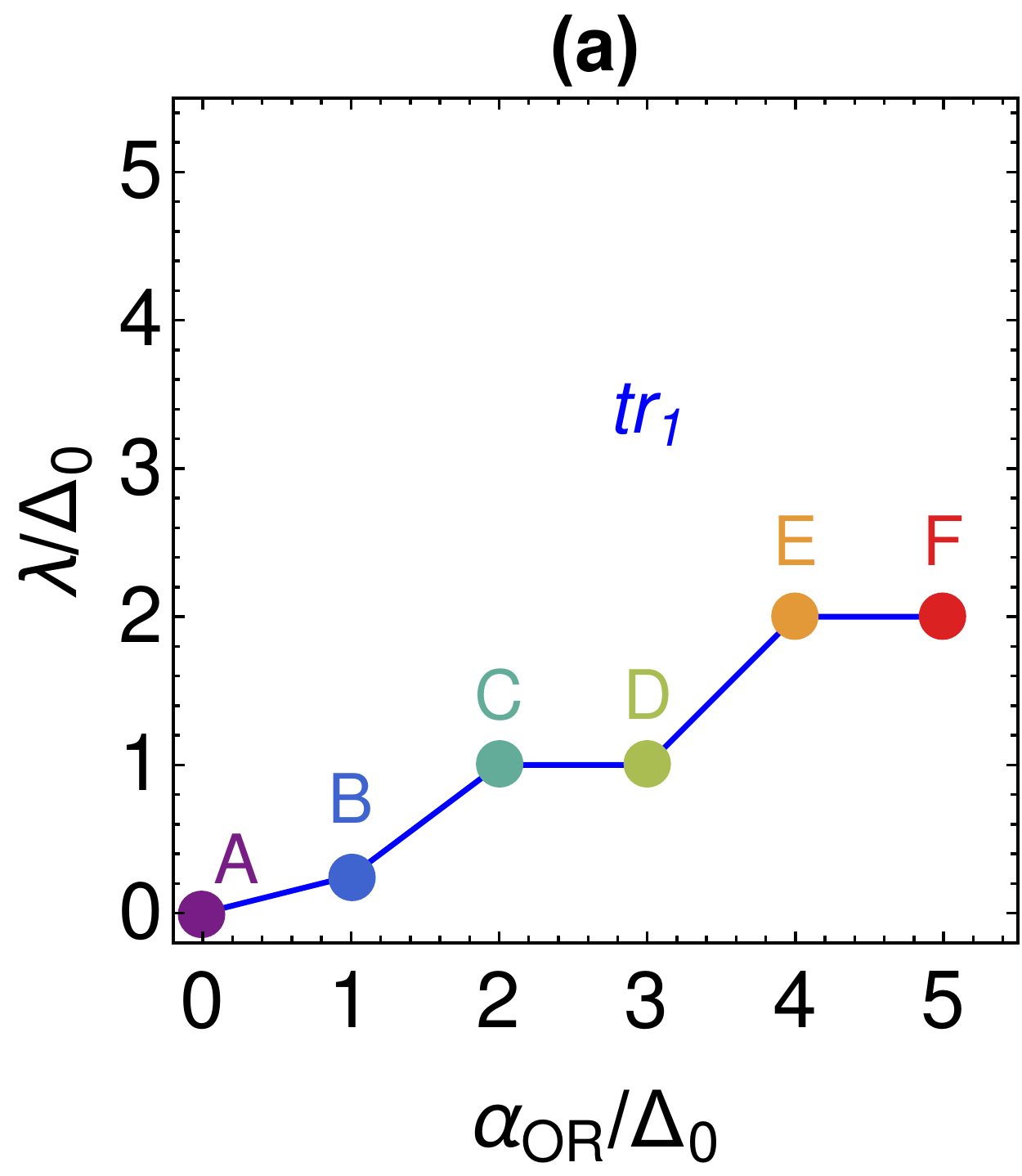} \hspace{0.1cm}
\includegraphics[width=0.25\textwidth]{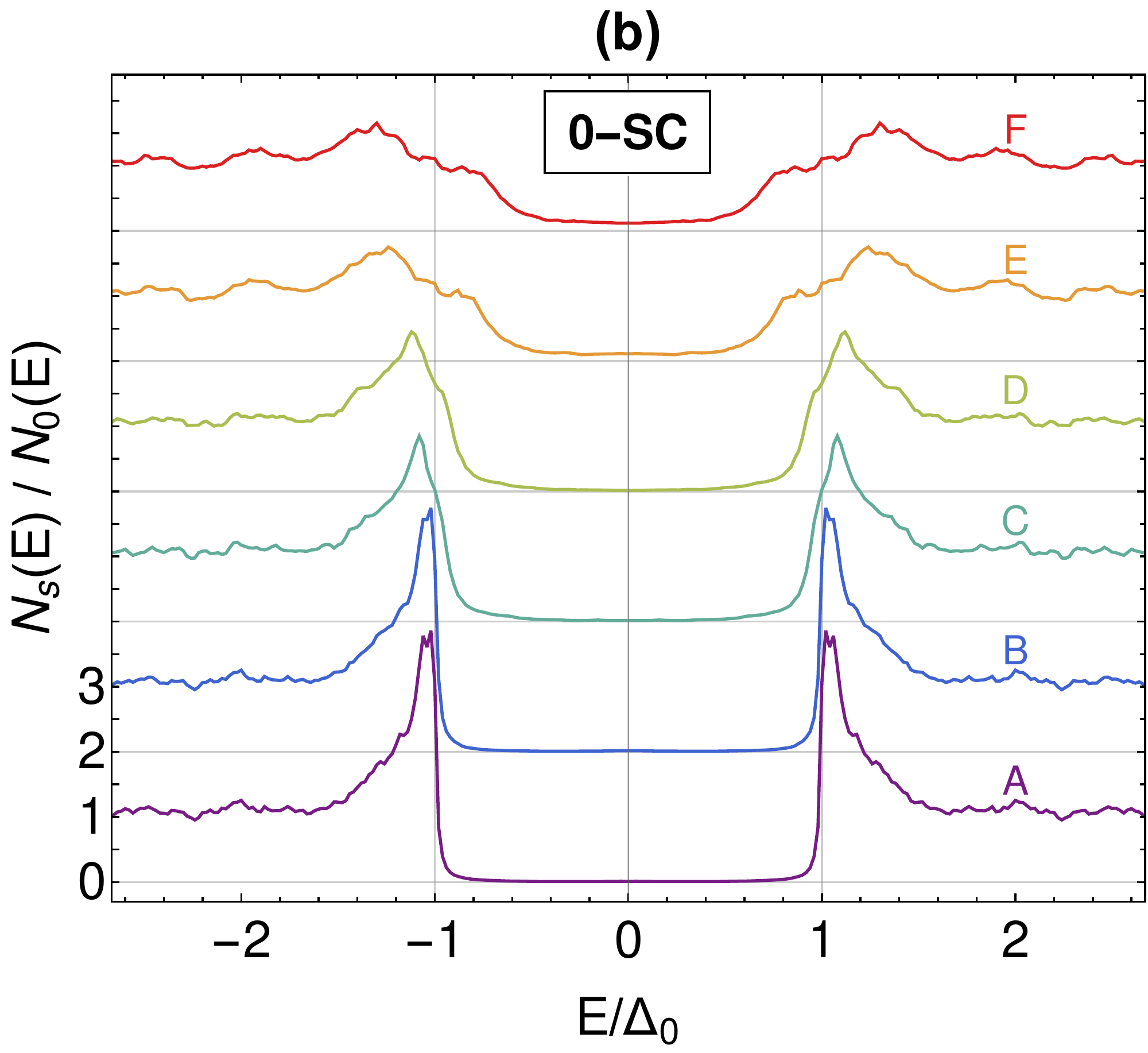} \hspace{0.08cm}
\includegraphics[width=0.25\textwidth]{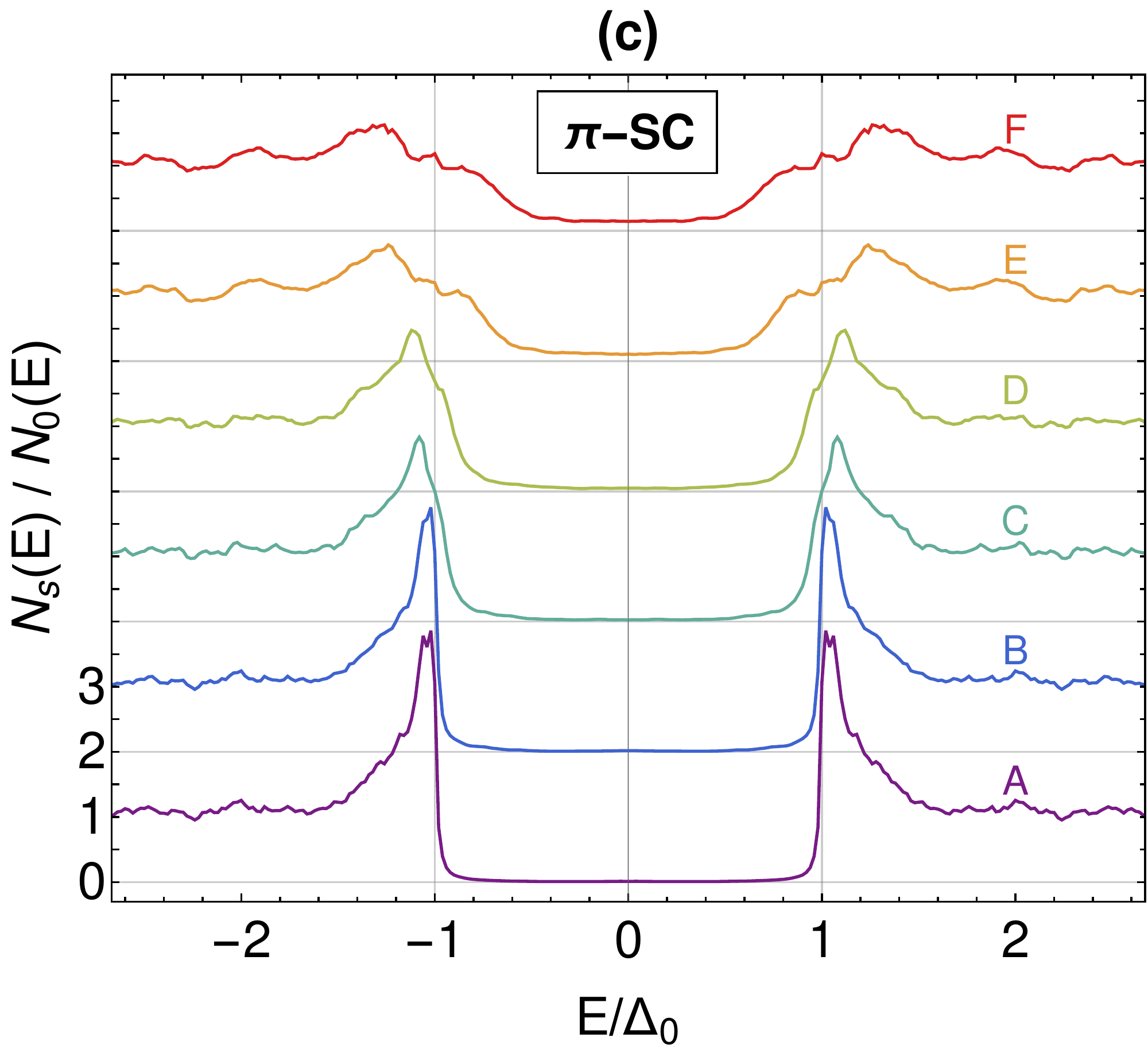} \hspace{0.08cm}
\includegraphics[width=0.25\textwidth]{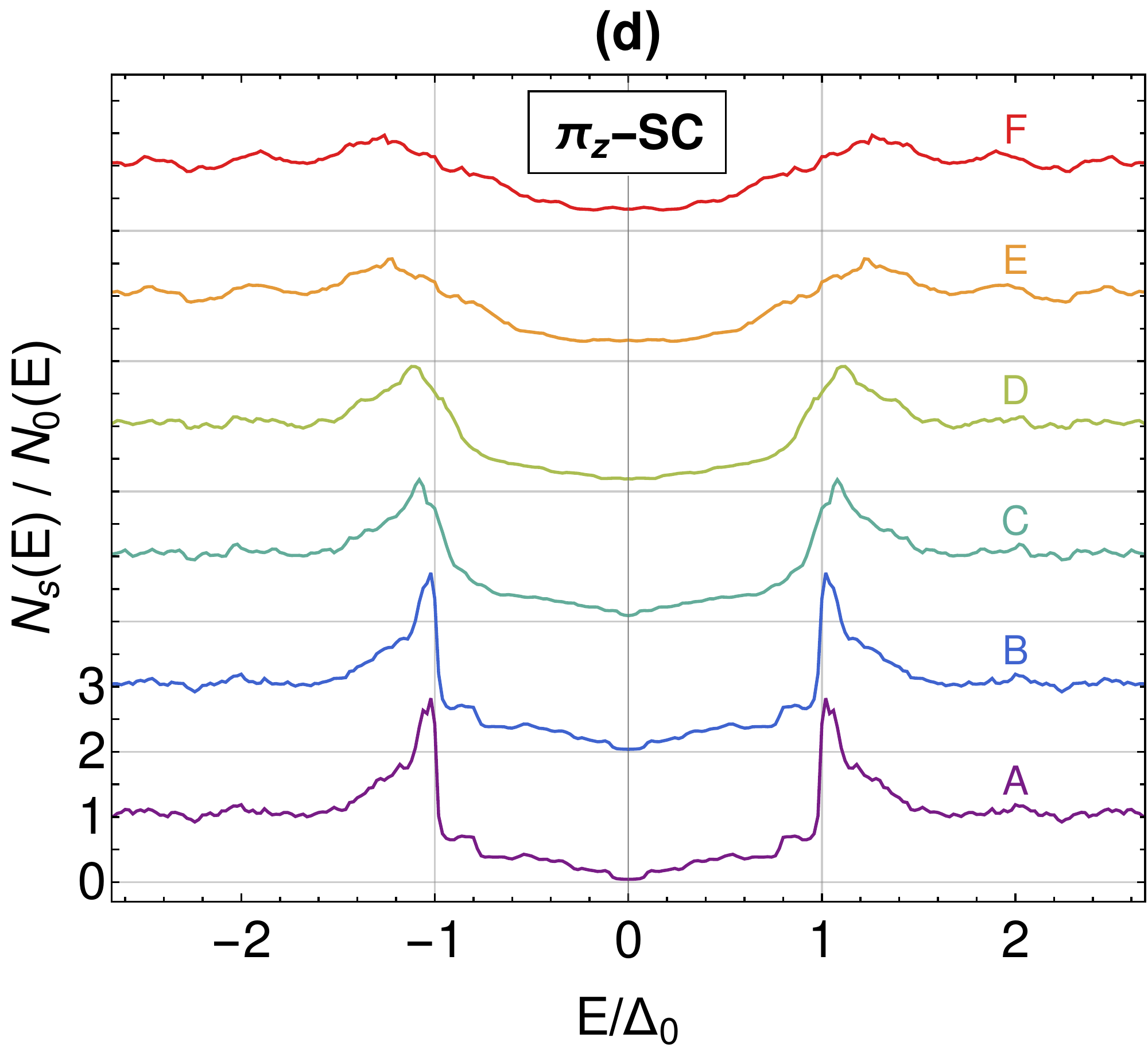} \\ \vspace{0.2cm}
\includegraphics[width=0.16\textwidth]{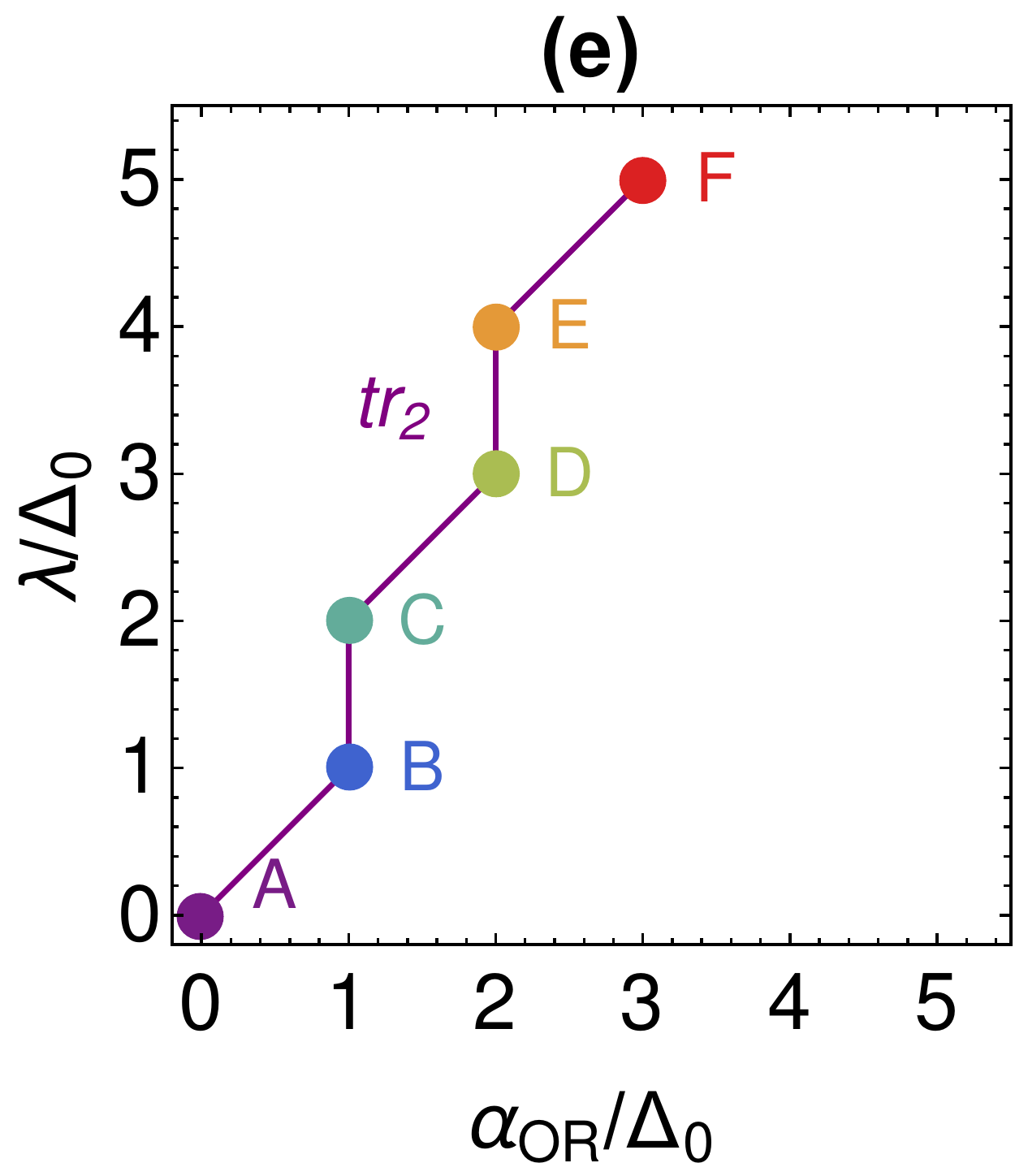} \hspace{0.1cm}
\includegraphics[width=0.25\textwidth]{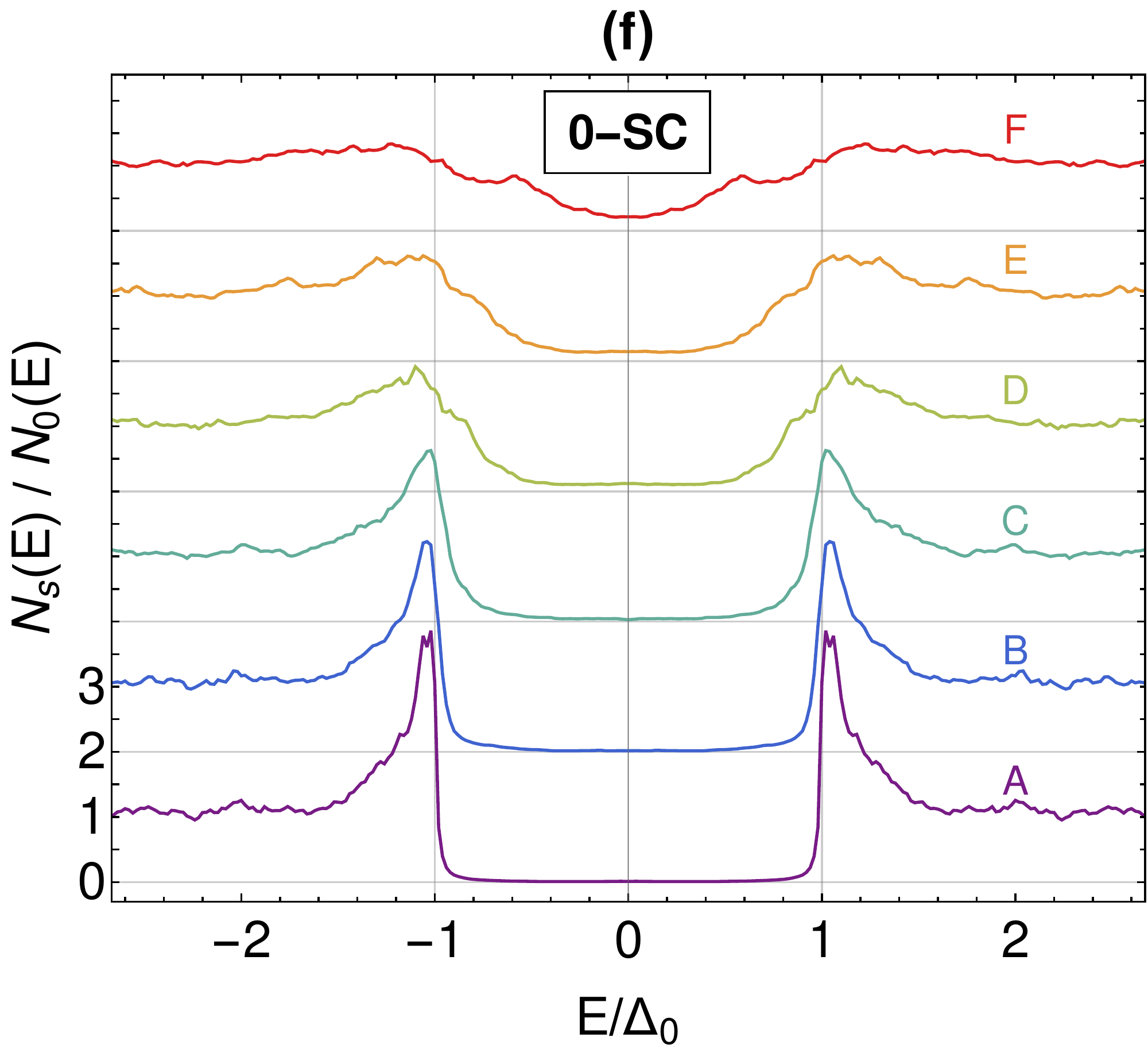} \hspace{0.08cm}
\includegraphics[width=0.25\textwidth]{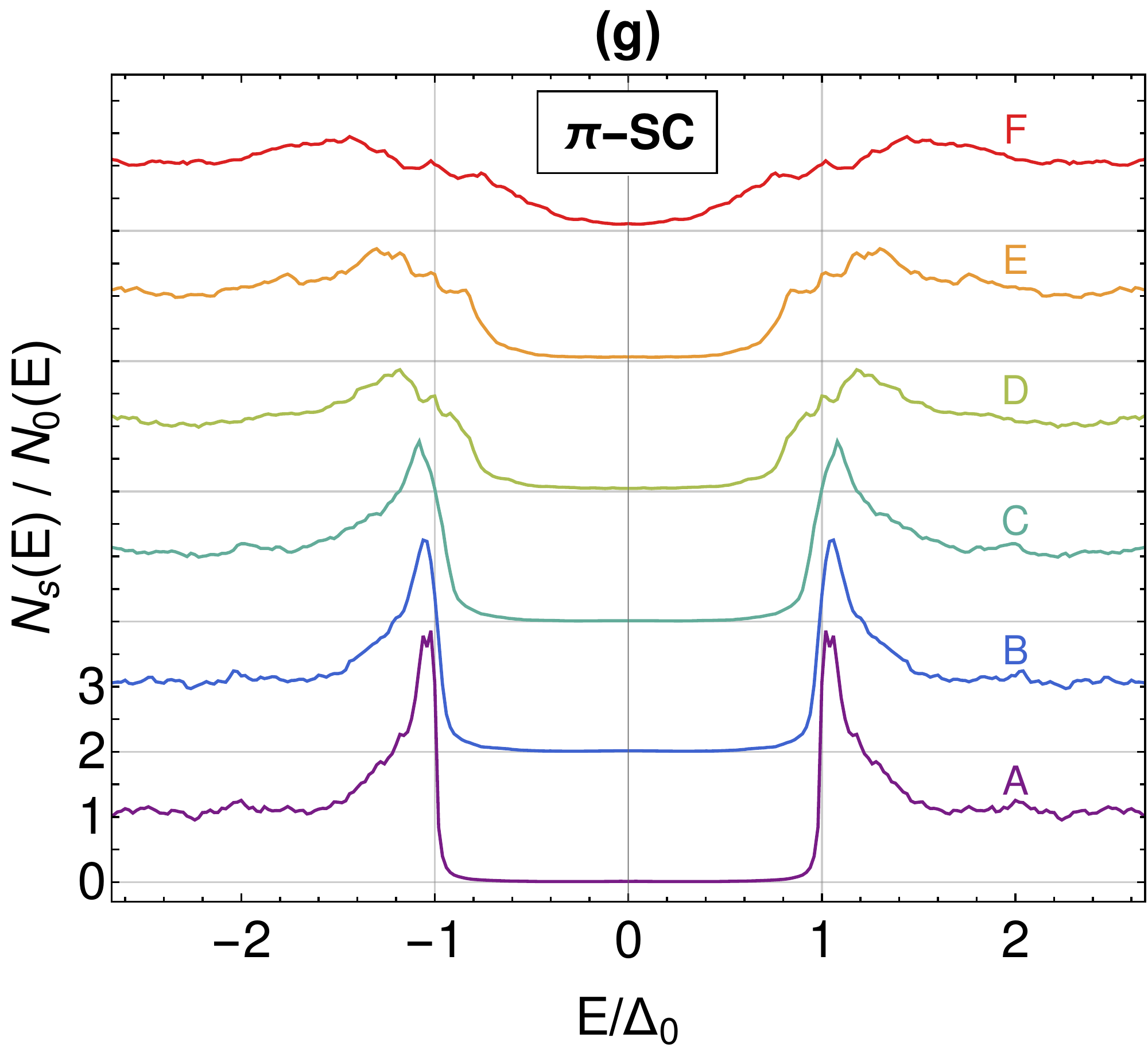} \hspace{0.08cm}
\includegraphics[width=0.25\textwidth]{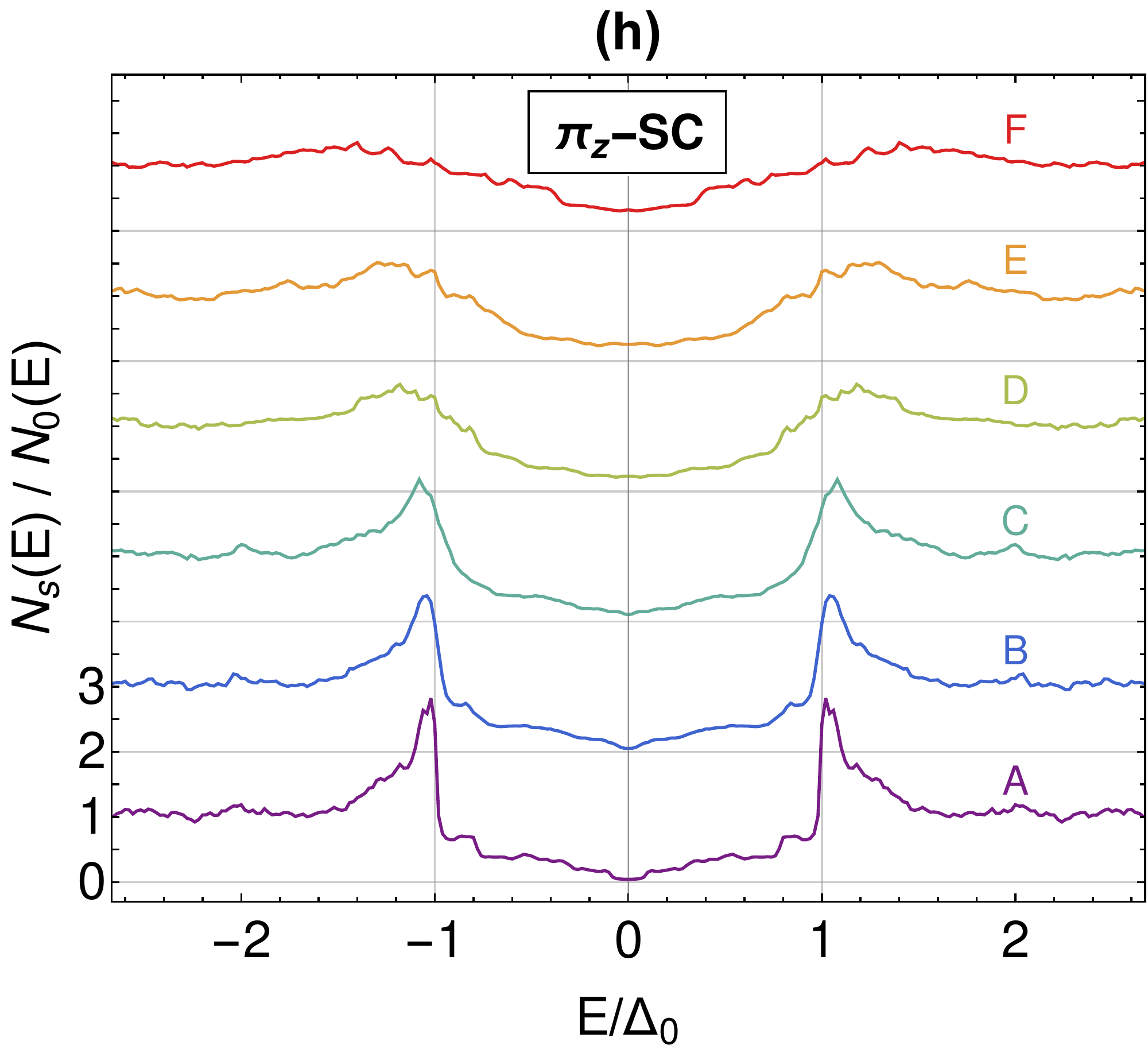}
\protect\caption{Zero temperature density of states of layered superconductor (with $n_z=6$ layers) subjected to an electric field through the modulation of the surface orbital Rashba couplings. For clarity the superconducting DOS ($N_s(E)$) is scaled to the normal state one ($N_n(E)$). The DOS from B to F are shifted along the vertical axis by 2. We investigate two representative paths in the $[\lambda,\alpha_{\text{OR}}]$ parameters space with (a) $\lambda<\alpha_{\text{OR}}$ ($tr_1$) and (e) $\lambda>\alpha_{\text{OR}}$ ($tr_2$), respectively. (b),(c) and (d) present the evolution of the density of states for the trajectory $tr_1$ in the 0-, $\pi$- and $\pi_z$- phases, respectively. The behavior of the density of states for the trajectory $tr_2$ is depicted in (f),(g) and (h) for the 0-,$\pi$- and $\pi_z$- phase, respectively. In-gap structures are more pronounced for the path $tr_2$. The suppression and shift of the peak at the gap edge is generally observed with an amplitude that depends on the character of the superconducting phase.}
\label{figDOS}
\end{figure*}
%%%%%%%%%%%%%

In Figs. \ref{SIS}(a) and (d) we present the evolution of the SIS$_2$ conductance for the 0-SC phase assuming that $\Delta_S=\Delta_0$ and $\Delta_{S_2}=2.5 \Delta_0$. This choice is convenient for having well separated characteristic voltages at the sum and difference of the superconducting gaps. At zero applied electric field the SIS tunneling conductance is basically marked by a main peak at $V_{+}=3.5 \Delta_0$ with satellite features at higher voltages oscillating around the normal state value. The variation of the surface orbital Rashba couplings along the paths with $\lambda$ greater or smaller than $\alpha_{{\text{OR}}}$ 
leads to distinct trends for the main peak. Indeed, for the trajectory with $\lambda<\alpha_{{\text{OR}}}$ (Fig. \ref{SIS}(a)) we observe a reduction of the intensity of the main conductance peak at $V_{+}$ that evolves into a dip accompanied by two peaks at voltages below and above $V_{+}$. On the other hand, when considering the $\lambda>\alpha_{{\text{OR}}}$ (Fig. \ref{SIS}(d)) we have that the reduction of the spectral weight of the peak at $V_{+}$ is accompanied by a downward shift of the conductance structures at low voltage bias. The result is a changeover  
from a single peak at $V_{+}$ into two structures, staying at $V_{+}$ and another one developing at $V_{*}\sim 3 \Delta_0$. The presence of the conductance peak at $V_{*}$ is connected with the bump structure developing in the DOS of the superconductor subjected to the electric field, at voltages inside the gap of the order of 0.5 $\Delta_0$ (Fig. \ref{figDOS}). We argue that the substantial filling up of the gap when $\lambda$ is larger than $\alpha_{{\text{OR}}}$ is the source of the peak formation in the SIS conductance at lower voltages with respect to $V_{+}$.
Furthermore, for this regime of coupling there is also a significant spectral weight redistribution at high voltages above $V_{+}$. 

Let us now move to the $\pi$-paired phases. As reported in Fig. \ref{SIS}(b),(e), for the uniform $\pi$-SC, we observe that the SIS conductance exhibits a profile having strong similarity with that of the 0-SC phase. Indeed, the variation of the electrically driven surface couplings suppresses the main peak at $V_{+}$ which evolves into multiple structures with a broad distribution with respect to the case without an applied electrostatic gating. The width of the conductance spectra is generally larger for the $\lambda>\alpha_{{\text{OR}}}$ trajectory, thus confirming the dominant role of the $\lambda$ coupling in yielding non-standard spectroscopic features. However, in this regime, differently from the 0-SC (Fig. \ref{SIS}(d)) we find that the distribution of the spectral weight keeps being symmetric around $V_{+}$. 

Finally, we discuss the SIS spectroscopic behavior assuming that one superconducting electrode is in the $\pi_z$ superconducting phase (Fig. \ref{SIS}(c),(f)). As we have understood from the SIN tunneling conductance, the $\pi_z$-SC is associated with a significant spectral weight in the gap. Then, already in the regime of small $\lambda$ and $\alpha_{{\text{OR}}}$, a non-vanishing conductance structure is obtained at voltage bias of $V_2=\Delta_{S_2}$. The resulting shoulder in the conductance spectra extends down to about $2 \Delta_0$ while moving along the trajectories in the phase space from $A$ to $F$ (Fig. \ref{SIS}(c),(f)). We also notice that the changeover of the conductance is not equivalent when comparing the paths with $\lambda<\alpha_{{\text{OR}}}$ or $\lambda>\alpha_{{\text{OR}}}$. 
In the former, the maximum of the conductance stays at $V_{+}$ while for the case with $\lambda>\alpha_{{\text{OR}}}$ the peak at $V_{+}$ becomes broad and flat. 

A relevant outcome of the above analysis is that although the main peak of the SIS conductance gets modified there are no traces of other structures emerging at a voltage bias that matches the difference of the superconducting gaps in the two electrodes. Furthermore, by a comparative inspection of the SIS tunneling conductance we can state that the main conductance peak gets generally suppressed by the application of an electrostatic gating resulting into a spectral redistribution with asymmetric features and broad structures whose profile is sensitive to microscopic details and to the character of the electrically induced superconducting phases. 

The observed features are directly linked to the modification of the DOS in the superconductor within the various induced phases (Fig. \ref{figDOS}). The quasiparticle peak at the gap edge typically shifts to high energies and gets suppressed in amplitude. The trajectory with $\lambda<\alpha_{{\text{OR}}}$ leads to an ingap structure below the gap edge that moves inwards within the gap accompanied with a slight increase of spectral weight as the strength of the electrostatic gating grows. In this regime, the main differences between the 0-,$\pi$- and $\pi_z$ phases occur close to zero energy. In fact, the $\pi_z$ configuration shows an enhanced filling up of the spectral weight in the gap due to the spatial sign change of the order parameter along the $z$-direction of the superconducting slab. 
As a consequence, one observes a non vanishing conductance already around $V_2=\Delta_{S_2}$ (Figs.5(c)-(f)) even for small values of $\alpha_{OR}$ and $\lambda$.
On the other hand, when considering the trajectory with $\lambda>\alpha_{{\text{OR}}}$ the suppression of the peak at the gap edge is generally more pronounced as well as the increase of the in gap spectral weight. When considering the $\pi$- and $\pi_z$-phases the peak at $E\sim \Delta_0$ loses its spectral weight and develops into multiple structures both at higher energies and inside the gap.

\section{Discussion and conclusions}

In conclusion we have unveiled the main spectroscopic features that would arise in SIN and SIS junctions with one electrode being a conventional superconductor subjected to an electrostatic gating that can drive the formation of orbital anti-phase paired phases by means of strong orbital Rashba effects at the surface. 
Our theoretical analysis reveals a complete set of distinctive marks for spectroscopically accessing the electrically induced superconducting phases and suggests an experimental way to disentangle thermal population unbalance from effects that are mainly due to the a variation of the electric field strength.

The analysis has been performed in a way to track the general evolution of the driven superconducting phases moving from the regimes of weak to strong gating amplitude. This approach is particularly useful in a system where the electrostatic field can lead to different types of transitions of the type 0-$\pi$ or 0-$\pi$-$\pi_z$ \cite{Mercaldo2020,Bours2020}. Here, the transitions are substantially accompanied by a phase rearrangement rather than an amplitude reconstruction of the band dependent order parameters. In this scenario, our results demonstrate that, since both SIN and SIS conductances share similar qualitative characteristics in the spectra while moving in the electric parameters phase space, we expect to observe smooth changeover across the transitions which would substantially manifest by a variation of the spectral weight in the conductance. 
This is an important outcome of the study because it tells that there will not be dramatic reconstructions of the conductance. On the contrary, one needs to track changes of the states in the gap or close to the conductance peak to identify the character of the electrically reconstructed superconducting states in the superconducting thin film. 
In particular, depending on whether the inter-layer orbital current processes are more relevant than the intra-layer ones the conductance both for the SIN and SIS can exhibit more pronounced features in the conductance. This is for instance the case of the shift to high voltages of the conductance maximum in the SIN configuration for the $\pi_z$ phase compared with 0-SC configuration which instead has a non-monotonic behavior.  

Another relevant concluding observation refers to the role of the thermal effects. We have demonstrated that the SIN conductance exhibits variations associated to a change of the electric field amplitude that can share wide similarities to those due to an increase of the temperature in the absence of an external electrical perturbation. This implies a difficulty to disentangle the two effects especially when the gating can also lead to heating or thermal gradient in the device. For this reason, we propose to combine the SIN with SIS tunneling probes. 
Our analysis indeed shows that at low temperature the electric field would not lead to the characteristic matching peak which is instead observed due to thermal excitations in conventional SIS junctions with different gap amplitudes. 
Finally, for the SIS configuration we have found that the application of an orbital polarizing electric field leads to a peculiar reconstruction of the main conductance peak with multiple structures whose profile is intricately tied to the character of the gate driven superconducting states. 

\begin{acknowledgments}
We acknowledge support by the EU’s Horizon 2020 research and innovation program under Grant Agreement nr. 964398 (SUPERGATE). 
F.G acknowledges the EU’s Horizon 2020 research and innovation program under Grant Agreement No. 800923 (SUPERTED) and the European Research Council under Grant Agreement No. 899315-TERASEC for partial financial support. M.C. acknowledges EU-H2020 Programme of QuantERA-NET Cofund in Quantum Technologies under Grant Agreement No. 731473 (QUANTOX). 
\end{acknowledgments}

\section{Appendix}

\subsection{Electrostatic potential close to the surface}

Concerning the charge reconstruction, in order to get an estimate of the induced electron density at the surface of the metal, we assume the electric field to be uniform in the $xy$ plane and the induced charge to have a peaked distribution, delta function, at the surface along the $z$- direction, i.e. $\rho(r)=\rho_S \delta(z)$. Then, for the electric field generated by the capacitor plates at a distance $L_C$ and the thin superconducting film of thickness $d$ placed inside the capacitor at equal distances from the two plates, there are two regions along the $z$ direction with inequivalent profile of the electrostatic potential $V(z)$. 
Region I for $-L_C/2<z<-d/2$ or $d/2<z<L_C/2$: we have $V(z)=\left( \frac{V_{G}}{L_C} \right) z-\frac{V_G}{2}$ or $V(z)=\left( \frac{V_{G}}{L_C} \right) z+\frac{V_G}{2}$. While for the region II inside the superconductor we have 
$V(z)=V_{II,m} \left[1+\tanh(-\lambda_{\text{TF}}(z+d/2) \right]$ with $V_{II,m}=- \frac{(d+L_C)V_G}{2 L}$ for $-d/2<z<0$ and $V(z)=V_{II,p} \left[1-\tanh(-\lambda_{\text{TF}}(z-d/2) \right]$ with $V_{II,p}=\frac{(d+L_C)V_G}{2 L}$ for $0<z<d/2$.
By integrating the Poisson equation close to the surface, i.e. at $z=-d/2$, one can obtain the expression for the induced in-plane charge density $\rho_S=4 \pi\epsilon_0 \left[ V_G/L_C + \lambda_{{TF}} \text{sech} (2 \lambda_{{TF}} d)^2 \right] $. For typical amplitudes of the applied voltages for the devices realized in Ref. \cite{DeSimoni2018} and related works, i.e. $V_G \sim 50 V$ with split gates placed at a distance $L_C \sim 100$ nm, we have that the variation of the electron density in the unit cell is of the order of $\delta n=0.05$ (assuming that the thickness $d=50$ nm, the in-plane unit cell size is $A_c=4 \AA^2$, and $\lambda_{{TF}}=4 \AA $ ). Such induced charge for a metal having an average density at the Fermi energy of $n_e\sim 1.0$ per unit cell, implies that the overall charge reconstruction due to the electrical gating is typically of the order of few percents. Since this variation is confined at the surface, in turn, it has also a small impact on the superconducting order parameter both at the surface and in the inner layers. 
\\
Similar conclusions on the negligible impact of the charge reconstruction in BCS metallic superconductors have been demonstrated in Refs. \cite{Virtanen2019,Chirolli2021}.

\subsection{Electrostatic potential and orbital Rashba coupling at the surface}

The external electric field on the surface of the superconductor is parallel to the out-of-plane $\hat{z}$-direction and thus the previously derived electrostatic potential close to the surface, at the linear order in $z$, can be described by a potential $V_{s}=-E_{s} z$ with $E_s$ being constant in amplitude (assuming the electric charge $e$ is unit). Hence, to construct the model Hamiltonian one has to consider a Bloch state representation and explicitly evaluate the matrix elements of the electrostatic potential $V_s$. 
Since the translational symmetry is broken along the $\hat{z}$-direction due to the finite thickness of the thin film and the presence of the electric field, the out-of-plane momentum, $k_z$, is not a good quantum number. Thus, a representation with a Bloch wave function associated to each layer is the most appropriate one to evaluate the effects of the electric field and the way it enters in the tight-binding model.
One can use the index $i_z$ to label different Bloch wave functions along the $\hat{z}$-direction as follows
\begin{eqnarray}
\psi_{{\bf k},\beta}({\bf r},i_z) =\frac{1}{\sqrt{N}}\sum_{\nu} \exp[i {\bf k} \cdot {\bf R}_{\nu, i_z}] \phi_{\beta}({\bf r}-{\bf R}_{\nu,i_z})
\end{eqnarray}
\noindent with the Bravais vector ${\bf R}_{\nu,i_z}$ identifying the position of the atoms in the $x-y$ plane for the layer labeled by $i_z$, $\beta$ indicating the atomic Wannier orbitals, and $N$ the total number of atomic sites. A central aspect in the derivation is that the atomic Wannier functions $\phi_{\beta}({\bf r}-{\bf R}_{\nu,i_z})$ span a manifold with non-vanishing angular momentum ${\bf L}$, e.g. $p$ or $d$ states. 
\\
The intra-layer matrix elements of the electrostatic potential can be then expressed as
\begin{eqnarray}
A^{||}_{l,m}=&& c_{\psi} \langle \psi_{{\bf k},l}({\bf r},i_z) | (-E_s z) | \psi_{{\bf k},m}({\bf r},i_z) \rangle \nonumber \\
=&&  c_{\psi} (-E_s) \frac{1}{N} \sum_{\nu,\gamma} \exp[i {\bf k} \cdot \left({\bf R}_{\nu, i_z}-{\bf R}_{\gamma, i_z}\right)] \times \nonumber \\
&& \times \int d^3 {\bf{r}} \phi^{*}_{l}({\bf r}-{\bf R}_{\nu,i_z})\,z\, \phi_{m}({\bf r}-{\bf R}_{\gamma,i_z}) \label{OR_coupling}\,
\end{eqnarray}
\noindent with $l$ and $m$ spanning the orbital space, and $c_{\psi}$ the normalization factor of the Bloch state.
Since the Wannier orbitals are significantly localized around the atomic position, the dominant terms are for nearest neighbor atoms. When evaluating those contributions in Eq. \ref{OR_coupling} for $l \neq m$ we obtain matrix elements that involve the $L_x$ and $L_y$ orbital angular momentum components and that are odd parity under in-plane spatial inversion \cite{Mercaldo2020}. The resulting term yields the intra-layer orbital Rashba coupling as in Eq. 2.
A similar derivation can be made for the inter-layer orbital Rashba term. 
%The inter-layer matrix element is in general complex because the electric field induces a time dependent vector potential along the $z$-direction that affects the relative phase of the Bloch functions in neighbor layers. This implies that one cannot fix the gauge in a way that the Bloch states in adjacent layers at the surface have the same phase. This is an overall phase factor that does not influence the amplitude of the coupling. 
The form of inter-layer term is due to the structure of the matrix elements of the electrostatic potential between Wannier functions in neighbor layers along the ${\hat z}-$direction. By evaluating these matrix elements \cite{Mercaldo2020}, it turns out that the electric field can induce an orbital polarization on nearest neighbors atoms in adjacent layers only if one allows for displacements/distortions of the atoms in the plane with respect to the high-symmetry positions. A detailed derivation of the interaction is presented in Ref. \cite{Mercaldo2020}.

\subsection{Temperature dependence of the superconducting order parameters}

\begin{figure*}[bt]
\includegraphics[width=0.93\textwidth]{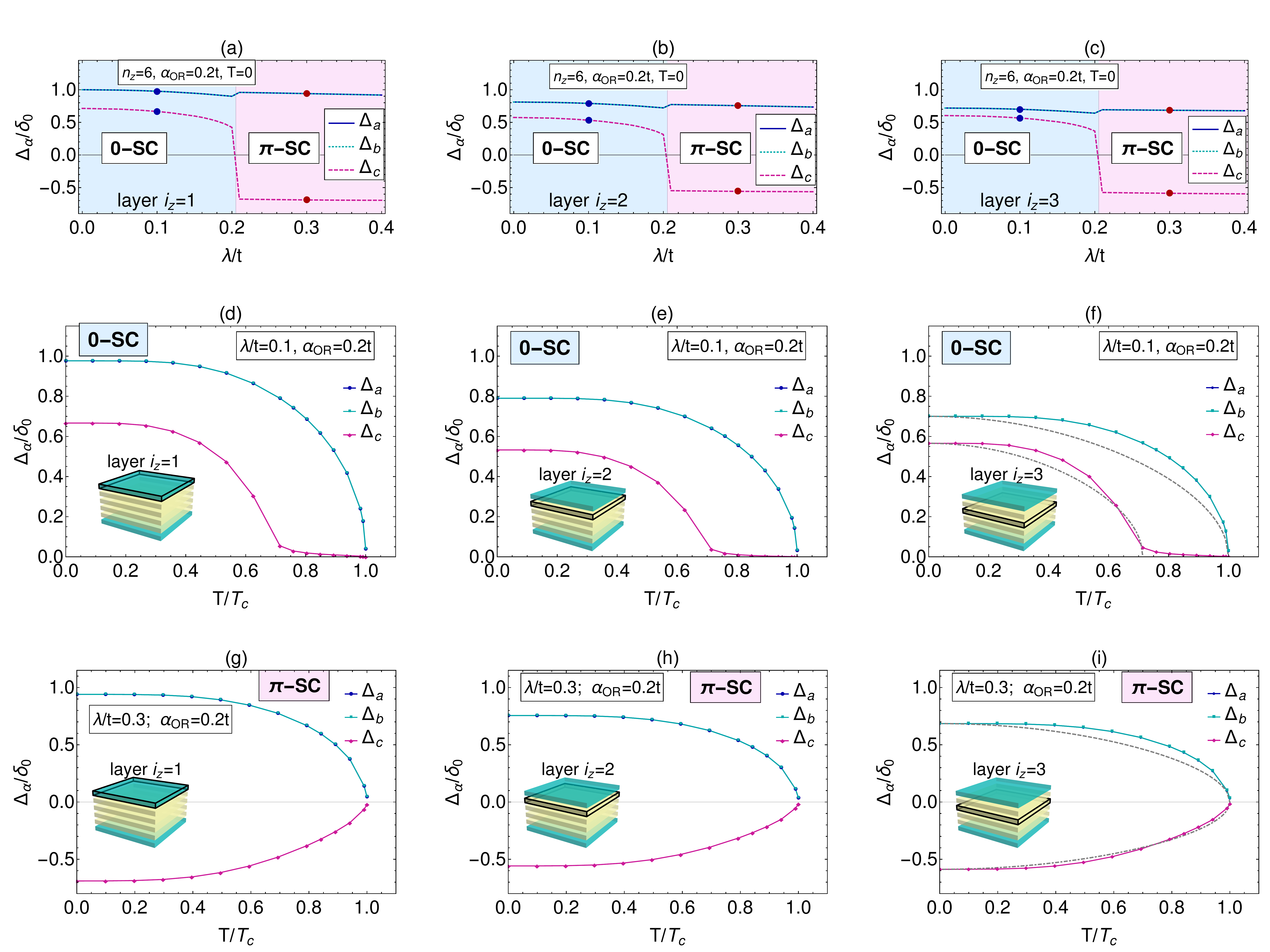} \hspace{0.4cm}
\protect\caption{{Behavior of the orbital dependent superconducting order parameter $\Delta_\alpha$ ($\alpha=a,b,c$), calculated with an iteratively self-consistent approach, %as a function of temperature 
for a system with $n_z=6$ layers, $t_\perp=1.5 t$, orbital Rashba coupling $\alpha_{OR}=0.2 t$. 
In (a)-(c) we report the superconducting order parameter at $T=0$ as a function of the inter-layer orbital Rashba coupling $\lambda$ focusing on the evolution of the the first three layers from the top surface (i.e. $i_z=1,2,3)$. The other corresponding layers starting from the bottom surface have the same amplitude. For the selected parameters we have a transition from the conventional (0-SC) to the $\pi-$phase ($\pi$-SC) for $\lambda=0.2t$. 
The dots indicate the values of $\lambda$ for which the orbital dependent superconducting OP have been evaluated as a function of the temperature in the panels (d)-(i).
In (d)-(f) we show the behavior of $\Delta(T/T_c)$ for the layers $i_z=1,2,3$, also marked in the insets, at $\lambda=0.1 t$.
In this regime, the superconductor exhibits a conventional SC phase (0-SC), with all $\Delta_\alpha$ having the same sign. 
In (g)-(h) the evolution $\Delta(T/T_c)$ is plotted for $\lambda=0.3\,t$, being an interaction amplitude that stabilizes an orbital anti-phase $\pi-$pairing state ($\pi$-SC), with a superconducting order parameter of a given orbital character having a $\pi$-shift in the phase respect to the other ones.
In panels (f) and (h) the gray lines indicate the temperature dependence of the canonical BCS gap as given by the phenomenological expression $\Delta(T)=\Delta_0 \cos(\sqrt{\pi T/T_c})$. Here, $\Delta_0$ is the value of $\Delta_\alpha(i_z,\lambda)$ at $T=0$. For the $c$ band in (f) to mimic the BCS profile for the smaller gap we assume an effective lower critical temperature.
In all panels $\delta_0$ is the value of $\Delta_a(\lambda=0)$ at the surface layer and it is used as a reference scale.
%We see that the OP of the $c$-band is suppressed at a lower temperature compared to the other bands.  
\label{fig_appendix_OP}
}}
\end{figure*}
\begin{figure}[bt]
\includegraphics[width=0.99\columnwidth]{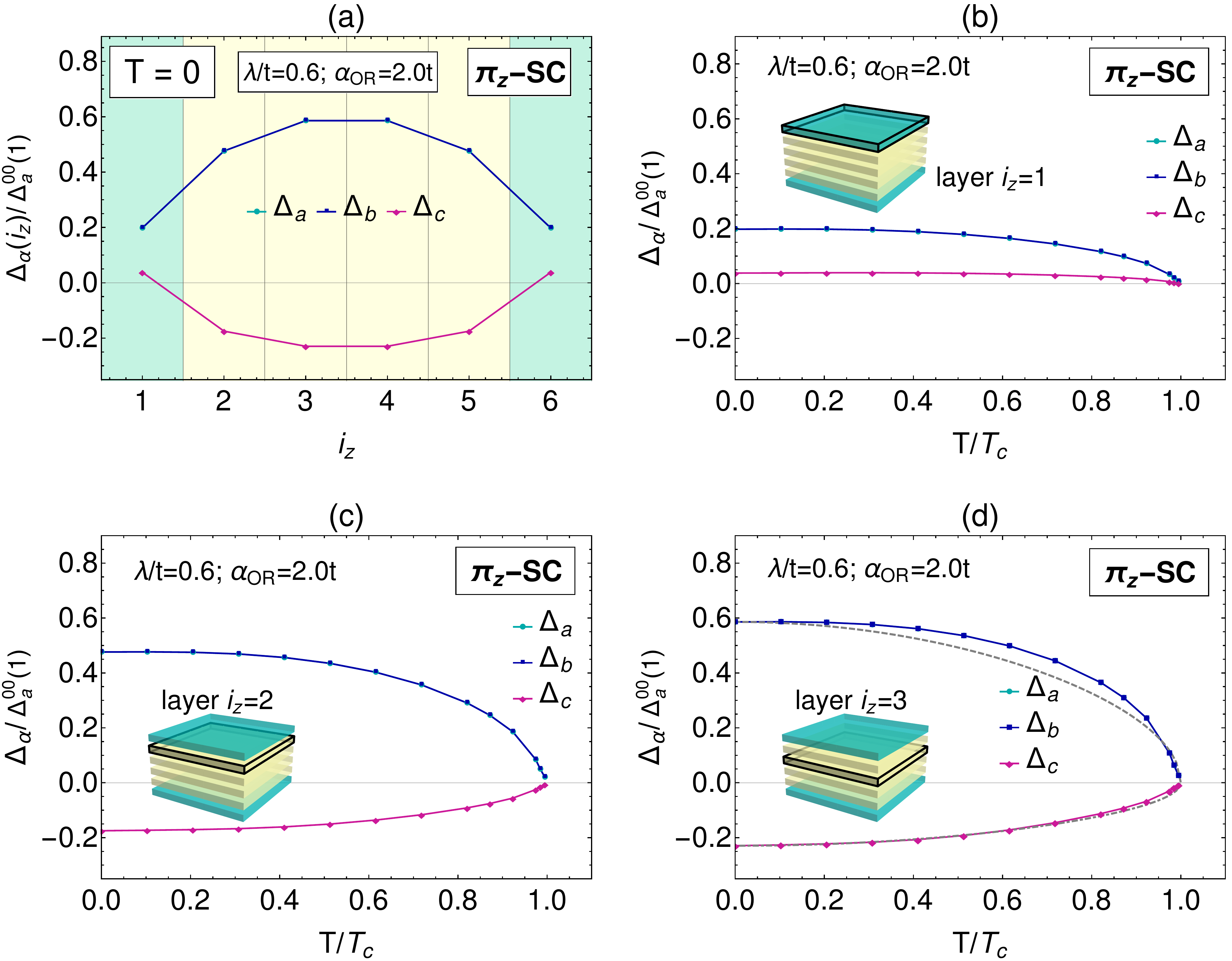} 
\protect\caption{{Behavior of the orbital dependent superconducting order parameter $\Delta_\alpha$ ($\alpha=a,b,c$), calculated with an iteratively self-consistent approach, %as a function of temperature 
for a system with $n_z=6$ layers and for a set of parameters such that the superconductor is in the $\pi_z$-phase, namely
 $t_\perp=0.9t$, orbital Rashba coupling $\alpha_{OR}=2.0t$ and inter-layer OR coupling $\lambda=0.6t$. 
In (a) we report the superconducting order parameter at $T=0$ as a function of the layer index $i_z$. 
In (b)-(d) we show the behavior of $\Delta(T/T_c)$ for the layers $i_z=1,2,3$, also marked in the insets. 
In fig. (d), the gray lines indicate the temperature dependence of the canonical BCS gap as given by the phenomenological expression $\Delta(T)=\Delta_0 \cos(\sqrt{\pi T/T_c})$, where  $\Delta_0$ is the value of $\Delta_\alpha(i_z,\lambda)$ at $T=0$. 
In all panels $\Delta_a^{00}(1)$ is the value of $\Delta_a(\alpha_{OR}=0,\lambda=0)$ at the surface layers and it is used as a reference scale in this figure.
%We see that the OP of the $c$-band is suppressed at a lower temperature compared to the other bands.  
\label{fig_appendix_OP_pz}
}}
\end{figure}

We have computed self-consistently the order parameters for each orbital component as a function of the temperature in order to compare it with the canonical BCS behavior and assess the possible consequences or deviations concerning the presented results.
The outcome is reported in Fig. \ref{fig_appendix_OP} for various representative cases in the phase space. As one can observe, the orbital dependence of the order parameters manifest in two aspects. The crystal field splitting and the orbital anisotropy of the kinetic energy can introduce an amplitude imbalance of the superconducting order parameters in the 0-SC at zero temperature with $\Delta_c$ being smaller than $\Delta_a$ and $\Delta_b$ (Fig. \ref{fig_appendix_OP} (a),(b),(c). This asymmetry evolves in temperature with the smaller gap typically having a suppression at temperatures in proximity of the superconducting transition $T_c$ (Fig. \ref{fig_appendix_OP} (d),(e),(f). As shown in Fig. \ref{fig_appendix_OP}, this behavior is substantially independent of the layer position. The resulting behavior in the 0-SC phase indicates that all the superconducting order parameters deviate from the canonical BCS profile (gray dotted line in Fig. \ref{fig_appendix_OP}(f)).
Moreover, the inter-layer orbital Rashba coupling modifies the tail of the small superconducting order parameter close to the critical temperature when considering the $\pi$-phase (Fig. \ref{fig_appendix_OP}(g),(h),(i)). In turn, for $\lambda$ values stabilizing the $\pi$-phase, we have that the temperature dependence of the superconducting order parameters follows quite well the BCS behavior (Fig. \ref{fig_appendix_OP}(i)). 
A similar trend is also obtained for the $\pi_z$-phase (Fig. \ref{fig_appendix_OP_pz}) where a good degree of matching with the BCS profile is obtained (Fig. \ref{fig_appendix_OP_pz}(d)). On the basis of these results, we conclude that the tunneling conductance evaluated by means of the BCS profile are suitable in the $\pi$- and $\pi_z$ phases for fully reproducing the temperature dependence of the multi-orbital superconducting order parameters. On the other hand, for the 0-SC configuration one can expect that quantitative changes might occur, especially close to the critical transition. We argue that, although these outcomes are directly related to a specific microscopic model, the inequivalent temperature dependence of the 0- and $\pi$- phases can manifest in a different evolution of the main conductance peaks for the SIN and SIS spectra in the corresponding phases.

\end{document}